\documentclass[aps,pra,english,twocolumn,showpacs,preprintnumbers,amsmath,amssymb,floatfix,superscriptaddress,longbibliography]{revtex4-1}
\usepackage[T1]{fontenc}
\usepackage[latin9]{inputenc}
\usepackage{graphicx}
\usepackage{amssymb}
\usepackage{babel}
\makeatletter
\usepackage[version=3]{mhchem}
\usepackage{color}

\global\arraycolsep=2pt
\usepackage{stmaryrd}
\usepackage{amsmath,braket}
\usepackage{amssymb}
\usepackage{graphicx}
\usepackage{textcomp}
\usepackage{calrsfs}
\usepackage{yfonts}
\usepackage{bm}
\usepackage{color}
\newcommand{\ben}{\begin{equation*}}
\newcommand{\een}{\end{equation*}}
\newcommand{\bean}{\begin{eqnarray*}}
\newcommand{\eean}{\end{eqnarray*}}

\newcommand{\be}{\begin{equation}}
\newcommand{\ee}{\end{equation}}
\newcommand{\bea}{\begin{eqnarray}}
\newcommand{\eea}{\end{eqnarray}}

\DeclareMathOperator{\tr}{tr}

\makeatother
\usepackage{babel}
\makeatother

\usepackage{color}

\usepackage{hyperref}
\hypersetup{
  colorlinks = true,
  citecolor  = blue,  
  linkcolor  = magenta 
}

\begin{document}
\title{Casimir stresses of the dielectric ball: inhomogeneity and divergences}

\author{Yang Li}
  \email{leon@ncu.edu.cn}
  \affiliation{School of Physics and Materials Science, Nanchang University, Nanchang 330031, China}
  \affiliation{Institute of Space Science and Technology, Nanchang University, Nanchang 330031, China}


\begin{abstract}
Puzzles are still preventing people from further understanding and manipulating the Casimir interaction in spherical systems. Here we investigate the behaviors of Casimir stresses in the system consisting of a ball immersed in the background, emphasising the roles of spherical geometry and inhomogeneity. Spherical modes are employed to evaluate the Green's dyadic and thus the Casimir stresses. The inhomogeneity of the media essentially modifies the wave form of the spherical mode, leading to significant impacts on the Casimir stresses, especially when far away from the surface of the ball. As the surface approached, the divergence (surface divergence) in Casimir stresses is seen. For both homogeneous and inhomogeneous cases, the leading behaviors (zero for the radial component, and inverse quantic order of distance for the transverse components) of Casimir stresses are exactly the same as those for the corresponding planar homogeneous wall, involving only properties of media at the surface and reflecting no information about the spherical geometry and the inhomogeneity, which implies the local nature. The other surface divergences are influenced by the spherical geometry, and for the transverse component always weaker than the planar contribution. The general impacts from the inhomogeneity of media to the surface divergences are also shown. The inhomogeneity will further soften the surface divergence. For two touching media with permittivities and permeabilities equal up to high enough order of their expansion over the distance to the surface, surface divergences may disappear together with the interaction Casimir stresses. Other factors, such as the refractivity and anisotropy, are also included, which may rise considerable complexities, but typically not related to divergences. Perspectives on the renormalization of the surface divergence are briefly outlined.
\end{abstract}

\date{\today}
\maketitle

\section*{Introduction}
\label{I}

\par Since Casimir~\cite{casimir1948attraction} predicted the observable effect of the zero-point energy of the quantum field in 1948, a large class of phenomena due to the nontrivial fluctuation of the ground state of the quantum field as well as their thermal corrections, are referred to as the Casimir effects in general (corresponding interaction is called the Casimir interaction, or van der Waals interaction in non-retarded cases), which have been explored and applied to diverse situations in practical scenarios and theoretical studies~\cite{milton2003casimir,bordag2009advances,Dalvit2011Casimir,milton2022state}. One of particular interest is about the Casimir stresses and energy in the spherical configuration.


\par On the early stage, interests on the Casimir interation in spherical geometry is due to a proposal by Casimir~\cite{casimir1953introductory} that the Casimir interaction could act as the Poincar\'{e} stress to balance the repulsive electrostatic force and lead us to a stable semi-classical electron model. Surprisingly, Boyer~\cite{boyer1968quantum} pointed out, and later justified again and again~\cite{balian1978electromagnetic,milton1978casimir,davies1972quantum}, that if the electron is modeled as a perfectly conducting spherical shell of zero thickness, the Casimir stress on this shell is repulsive, in contrast to Casimir's original plate model where the Casimir force is attractive. This result not only casts serious doubts on the validity of Casimir's semi-classical model for the electron, but also illustrates the highly non-trivial dependence of Casimir interaction on the geometry of the system. When extending the shell model to the dielectric ball, the difficulties immediately rise because of the divergences in the Casimir self-energy of the ball~\cite{milton1980semiclassical}. It was recognized in the isorefractive cases, these divergences can be canceled out~\cite{brevik1982casimir,brevik1988casimir}. But generally these divergences can not be systematically renormalized, and ambiguities are thus raised, hindering the study on some real-world problems. For instance, it was proposed by Schwinger~\cite{schwinger1993casimir,schwinger1994casimir} that the Casimir self-energy of the bubble can play a role in the sonoluminescence. Although Milton et. al.~\cite{milton1997casimir} has claimed that up to the leading order, the Casimir self-energy is too small to contribute significantly in the sonoluminescence process, the ambiguity of the self-energy due to the divergences in its full expression still prevents us from drawing a clear-cut conclusion. In recent years, researchers are still taking efforts to put these divergences under control. Milton and Brevik~\cite{milton2018casimir} tried to combine the isorefractivity and zero-thickness of the shell, but only got a partial success. Leonhardt et. al.~\cite{avni2018casimir} claimed they found the way out, but was realized do not deal with the bulk divergences properly~\cite{milton2020self}.

\par Moreover, extra complexities are introduced when the inhomogeneity of media taken into account. For example, Simpson et. al.~\cite{leonhardt2011exact} tried to show an attractive Casimir force in a spherical configuration consisting of an inhomogeneous ball, namely the Maxwell's fish eye, within a perfect conductor. In this paper, we would like to further explore the profound impacts of the inhomogeneity of media on the divergent behaviors of the Casimir stresses (Casimir energy implied) in the spherical configuration with the spherical symmetry preserved. Casimir stresses in the system consisting of a dielectric ball immersed in the background are investigated. These Casimir stresses will be divided into the bulk and interaction parts, and we mainly focus on the interaction contributions, since the bulk contributions, though always divergent due to the self-interaction, are relatively trivial and only briefly evaluated and discussed in the Appendix~\ref{ApB}.

\par This paper is organized as follows: In Sec.~I, the theory is provided in detail, based on which we in Sec.~II evaluate and analyze the Casimir stresses, both analytically and numerically, for homogeneous case and the inhomogeneous cases. In the close vicinity of the surface of the ball, the Casimir stresses behaves as their planar counterparts, while the influences of the spherical geometry can never be ignored, implying the global aspect of the Casimir interaction. The behaviors for general inhomogeneous cases near the surface are demonstrated, and the properties of the softening of surface divergences associated with the inhomogeneity are clarified. Arguments on the condition about the well-defined surface are accordingly given. Finally the conclusions are provided in Sec.~III. A brief evaluation on Casimir stresses in the relevant planar configuration is provided in Appendix~\ref{ApA}.

%
%

\par The natural unit $\hbar=c=\varepsilon_0=\mu_0=1$ is used, unless otherwise specified.

\section{Theory}
\label{T}
\par It is well-know that the momentum of the electromagnetic field should satisfies the relation as in Ref.~\cite{schwinger2000classical}
\begin{equation}
\label{eq.or}
\mathbf{f}+\frac{\partial\mathbf{G}}{\partial t}=\nabla\cdot(\mathbf{DE}+\mathbf{BH})-\nabla_E(\mathbf{D}\cdot\mathbf{E})-\nabla_H(\mathbf{B}\cdot\mathbf{H})
,
\end{equation}
where $\nabla_{E(H)}$ means the operator only acts on $\mathbf{E}$ ($\mathbf{H}$), $\mathbf{f}=\rho\mathbf{E}+\mathbf{j}\times\mathbf{B}$ with $\rho$ and $\mathbf{j}$ being the free charge and current densities, and the momentum density of the field is $\mathbf{G}=\mathbf{D}\times\mathbf{B}$. Consider the time-averaging of Eq.~\eqref{eq.or} over the time $T$, which is macroscopically small but microscopically large, then with the following Fourier forms for vectors and tensors, using the electric field $\mathbf{E}(t)$ and the permittivity $\bm{\varepsilon}(t)$ as examples respectively,
\begin{equation}
\mathbf{E}(t)=\int\frac{d\omega}{\sqrt{2\pi}}\mathbf{E}(\omega)e^{-i\omega t},\ \bm{\varepsilon}(t)=\int\frac{d\omega}{2\pi}\bm{\varepsilon}(\omega)e^{-i\omega t},
\end{equation}
then the righ side of Eq.~\eqref{eq.or} takes the averaged form
\begin{eqnarray}
\label{eq.or3}
& &
\frac{1}{T}\int d\omega\bigg\{\nabla\cdot\bigg[\mathbf{D}(\omega)\mathbf{E}(-\omega)
+\mathbf{B}(\omega)\mathbf{H}(-\omega)\bigg]
\nonumber\\
& &
-
\nabla_E\tr\mathbf{D}(\omega)\mathbf{E}(-\omega)
-
\nabla_H\tr\mathbf{B}(\omega)\mathbf{H}(-\omega)\bigg\},
\nonumber\\
&=&
\frac{1}{T}\int d\omega\bigg\{\nabla\cdot\bigg[\bm{\varepsilon}(\omega)\cdot\mathbf{E}(\omega)\mathbf{E}(-\omega)
\nonumber\\
& &
+
\bm{\mu}(\omega)\cdot\mathbf{H}(\omega)\mathbf{H}(-\omega)\bigg]
-
\nabla_E\tr\bm{\varepsilon}(\omega)\cdot\mathbf{E}(\omega)\mathbf{E}(-\omega)
\nonumber\\
& &
-
\nabla_H\tr\bm{\mu}(\omega)\cdot\mathbf{H}(\omega)\mathbf{H}(-\omega)\bigg\},
\end{eqnarray}
where the constitutive relations below are included
\begin{eqnarray}
\mathbf{D}(\omega)=\bm{\varepsilon}(\omega)\cdot\mathbf{E}(\omega),\ \mathbf{B}(\omega)=\bm{\mu}(\omega)\cdot\mathbf{H}(\omega).
\end{eqnarray}
Here we ignore the dissipation and assume the Hermiticity of the media, for simplicity and clarity as did in Ref.~\cite{parashar2018quantum}, that is, $\bm{\varepsilon}^{\dag}(\omega)=\bm{\varepsilon}(-\omega)$ and $\bm{\mu}^{\dag}(\omega)=\bm{\mu}(-\omega)$. Then the last two terms in Eq.~\eqref{eq.or3} can be written as
\begin{eqnarray}
& &
\frac{1}{T}\int d\omega\bigg[
\nabla\frac{\tr\bm{\varepsilon}(\omega)\cdot\mathbf{E}(\omega)\mathbf{E}(-\omega)
+
\tr\bm{\mu}(\omega)\cdot\mathbf{H}(\omega)\mathbf{H}(-\omega)}{-2}
\nonumber\\
& &
+
\nabla_{\varepsilon}\frac{\tr\bm{\varepsilon}(\omega)\cdot\mathbf{E}(\omega)\mathbf{E}(-\omega)}{2}
+
\nabla_{\mu}\frac{\tr\bm{\mu}(\omega)\cdot\mathbf{H}(\omega)\mathbf{H}(-\omega)}{2}\bigg],
\nonumber\\
&&
\end{eqnarray}
with $\nabla_{\varepsilon(\mu)}$ acting only on $\bm{\varepsilon}$ ($\bm{\mu}$). When the medium is homogeneous, Eq.~\eqref{eq.or3} is simply $-\nabla\cdot\overline{\mathbf{T}}$, and the electromagnetic tensor $\mathbf{T}$ of the field-medium system is symbolically
\begin{eqnarray}
\mathbf{T}
&=&
\frac{\bm{1}}{2}\tr(\bm{\varepsilon}\cdot\mathbf{EE}+\bm{\mu}\cdot\mathbf{HH})
-\bm{\varepsilon}\cdot\mathbf{EE}-\bm{\mu}\cdot\mathbf{HH},
\end{eqnarray}
implying the local conservation of momentum. On the other hand, if the medium is inhomogeneous, Eq.~\eqref{eq.or3} gains an extra contribution, i.e., $-\nabla\cdot\overline{\mathbf{T}}+\overline{\bm{\phi}}$, with $\bm{\phi}$ being
\begin{eqnarray}
\bm{\phi}
&=&
\frac{1}{2}\nabla_{\varepsilon}\tr\bm{\varepsilon}\cdot\mathbf{EE}
+\frac{1}{2}\nabla_{\mu}\tr\bm{\mu}\cdot\mathbf{HH}.
\end{eqnarray}
As for the energy of the electromagnetic field, similar arguments follows. The energy relation is~\cite{schwinger2000classical}
\begin{equation}
\mathbf{j}\cdot\mathbf{E}+\nabla\cdot(\mathbf{E}\times\mathbf{H})=-\mathbf{E}\cdot\frac{\partial\mathbf{D}}{\partial t}-\mathbf{H}\cdot\frac{\partial\mathbf{B}}{\partial t}
,
\end{equation}
with $\mathbf{j}$ being the free current density, which, in the time-averaging as above, leads us to the energy density of the system expressed as
\begin{eqnarray}
\overline{U}
&=&
\int \frac{d\omega}{2T}\tr\bigg[
\frac{\partial\omega\bm{\varepsilon}(\omega)}{\partial \omega}\cdot\mathbf{E}(\omega)\mathbf{E}(-\omega)
\nonumber\\
& &
+
\frac{\partial\omega\bm{\mu}(\omega)}{\partial \omega}\cdot\mathbf{H}(\omega)\mathbf{H}(-\omega)
\bigg].
\end{eqnarray}

\par To investigate the stress tensor and the energy density induced by the quantum fluctuation of the field on the ground state, namely the Casimir stress tensor and energy density, the quantum averaging should be taken into account. It is convenient to evaluate those relevant quantities in the Euclidean space with the imaginary frequency $\omega\rightarrow i\zeta$. The vacuum correlation of the electric field can be written in terms of the Green's dyadic~\cite{milton2010casimir,parashar2018quantum,milton2020self}
\begin{eqnarray}
\braket{\mathbf{E}(\zeta,\mathbf{r})\mathbf{E}(\zeta',\mathbf{r}')}
=-\delta(\zeta+\zeta')\bm{\Gamma}_{\zeta}(\mathbf{r},\mathbf{r}'),
\end{eqnarray}
in which the the Green's dyadic $\bm{\Gamma}_{\zeta}(\mathbf{r},\mathbf{r}')$ is defined with the equation
\begin{equation}
\label{eq.gam}
\bigg[\bm{\varepsilon}(\zeta,\mathbf{r})+\frac{\nabla\times\bm{\mu}^{-1}(\zeta,\mathbf{r})
\cdot\nabla\times\bm{1}}{\zeta^2}\bigg]\cdot\bm{\Gamma}_{\zeta}(\mathbf{r},\mathbf{r}')
=\bm{1}\delta(\mathbf{r}-\mathbf{r}').
\end{equation}
For the magnetic field, the correlation has a similar form, that is,
\begin{eqnarray}
\braket{\mathbf{H}(\zeta,\mathbf{r})\mathbf{H}(\zeta',\mathbf{r}')}
=-\delta(\zeta+\zeta')\bm{\Phi}_{\zeta}(\mathbf{r},\mathbf{r}'),
\end{eqnarray}
where $\bm{\Phi}_{\zeta}(\mathbf{r},\mathbf{r}')$ satisfies a similar equation as  $\bm{\Gamma}_{\zeta}(\mathbf{r},\mathbf{r}')$, except for a substitution $\bm{\varepsilon}\leftrightarrow\bm{\mu}$. The vacuum expectation values of the stress tensor $\overline{\mathbf{T}}$ and energy density $\overline{u}$ are thus expressed as
\begin{eqnarray}
\overline{\mathbf{T}}(\mathbf{r})
&=&
\int\frac{d\zeta}{2\pi}\bigg[
\frac{\tr\bm{\varepsilon}(\zeta,\mathbf{r})\cdot\bm{\Gamma}_{\zeta}(\mathbf{r},\mathbf{r})
+
\tr\bm{\mu}(\zeta,\mathbf{r})\cdot\bm{\Phi}_{\zeta}(\mathbf{r},\mathbf{r})}{-2}\bm{1}
\nonumber\\
& &
+\bm{\varepsilon}(\zeta,\mathbf{r})\cdot\bm{\Gamma}_{\zeta}(\mathbf{r},\mathbf{r})
+\bm{\mu}(\zeta,\mathbf{r})\cdot\bm{\Phi}_{\zeta}(\mathbf{r},\mathbf{r})\bigg],
\end{eqnarray}
\begin{eqnarray}
\overline{U}(\mathbf{r})
&=&
-\int\frac{d\zeta}{4\pi}\tr\bigg\{
\frac{\partial[\zeta\bm{\varepsilon}(\zeta,\mathbf{r})]}{\partial\zeta}
\cdot\bm{\Gamma}_{\zeta}(\mathbf{r},\mathbf{r})
\nonumber\\
& &
+
\frac{\partial[\zeta\bm{\mu}(\zeta,\mathbf{r})]}{\partial\zeta}\cdot\bm{\Phi}_{\zeta}(\mathbf{r},\mathbf{r})
\bigg\}.
\end{eqnarray}
The averaging time period is $T=2\pi\delta(0)$ here. Therefore, the explicit expressions for the Green's dyadics are required to analyze behaviors of the Casimir stress tensor and energy density.

\par According to Eq.~\eqref{eq.gam} and that for $\bm{\Phi}$, the permittivity and permeability impact the correlation functions significantly. Generally, the analytic formulas for $\bm{\Gamma}$ and $\bm{\Phi}$ are, if available any way, pretty complicated. In this paper, we focus on the spherical configuration consisting of a dielectric ball in the background with its center located at the origin without losing any generality. The spherical symmetry of this system is maintained for simplicity, that is, the system is symmetry under the rotation with respect to the center of the ball. The responses of any medium in the configuration should depend only on the radius due to the symmetry, and we, in the aim of studying the influences of anisotropy, take the permittivity in the form
\begin{eqnarray}
\label{eq.spepsilon}
\bm{\varepsilon}(\zeta,\mathbf{r})=(\bm{1}-\mathbf{\hat{r}\hat{r}})\varepsilon_t(\zeta,r)
+\mathbf{\hat{r}\hat{r}}\varepsilon_p(\zeta,r),
\end{eqnarray}
where $\hat{\mathbf{r}}$ is the unit vector in the radial direction. The same notation applies to the  permeability except for $\varepsilon$ replaced by $\mu$. Here transverse components (i.e. the polar and and azimuthal components) of the permittivity and permeability are set equal and could deviate from their radial part. Then by separating the Green's dyadics into contributions from transverse electric (TE) and transverse magnetic (TM) modes, they can be expressed in terms of two scalar Green's functions defined below
\begin{subequations}
\label{eq.g}
\begin{equation}
\label{eq.ge}
\bigg[r\frac{d}{dr}\mu_t^{-1}\frac{d}{dr}r-\frac{l(l+1)}{\mu_p}-\varepsilon_t\zeta^2r^2\bigg]g^{\rm E}_{\zeta,l}(r,r')=\delta(r-r'),
\end{equation}
\begin{equation}
\label{eq.gh}
\bigg[r\frac{d}{dr}\varepsilon_t^{-1}\frac{d}{dr}r-\frac{l(l+1)}{\varepsilon_p}-\mu_t\zeta^2r^2\bigg]g^{\rm H}_{\zeta,l}(r,r')=\delta(r-r').
\end{equation}
\end{subequations}
Utilize the expansion based on the vector spherical harmonics, the Green's dyadic $\bm{\Gamma}$ can be written
\begin{eqnarray}
\label{eq.spGamma1}
\bm{\Gamma}_{\zeta}(\mathbf{r},\mathbf{r}')=\sum_{l=1}^{\infty}\sum_{m=-l}^l\sum_{i,j=1}^3g_{i,j}(r,r')
\mathbf{X}_{l,i}^m(\Omega)\mathbf{X}_{l,j}^{m*}(\Omega'),\quad\quad
\end{eqnarray}
where $\mathbf{X}_{l,i}^m(\Omega)$ signifies the vector spherical harmonics used here as follows
\begin{eqnarray}
\mathbf{X}_{l,1}^m=Y_l^m\hat{\mathbf{r}},\ \mathbf{X}_{l,2}^m=\frac{r\nabla Y_l^m}{\sqrt{l(l+1)}},\ \mathbf{X}_{l,3}^m=\frac{\mathbf{r}\times\nabla Y_l^m}{\sqrt{l(l+1)}}.\quad
\end{eqnarray}
Write out the explicit formula for $g_{i,j}$ with $g^{\rm E(H)}$, and $\bm{\Gamma}$ as a matrix with the element $g_{i,j}$, we arrive at
\begin{widetext}
\begin{eqnarray}
\label{eq.spGamma2}
\bm{\Gamma}_{\zeta}(\mathbf{r},\mathbf{r}')=\sum_{l=1}^{\infty}\sum_{m=-l}^l
\left[
  \begin{array}{ccc}
    \frac{l(l+1)}{\varepsilon_p\varepsilon'_prr'}g^{\rm H}_{\zeta,l} & \frac{\sqrt{l(l+1)}}{\varepsilon_p\varepsilon'_trr'}\frac{\partial(r'g^{\rm H}_{\zeta,l})}{\partial r'} &  \\
    \frac{\sqrt{l(l+1)}}{\varepsilon_t\varepsilon'_prr'}\frac{\partial(rg^{\rm H}_{\zeta,l})}{\partial r} & \frac{1}{\varepsilon_t\varepsilon'_trr'}\frac{\partial^2(rr'g^{\rm H}_{\zeta,l})}{\partial r\partial r'} &  \\
     &  & -\zeta^2g^{\rm E}_{\zeta,l} \\
  \end{array}
\right]
,
\end{eqnarray}
\end{widetext}
and $\bm{\Phi}_{\zeta}(\mathbf{r},\mathbf{r}')$ can be obtained with the substitutions $\varepsilon\leftrightarrow\mu$ and ${\rm E}\leftrightarrow{\rm H}$. Therefore, the difficulties lie on solving Eqs.~\eqref{eq.ge} and \eqref{eq.gh}, which can be further reduced to find the solutions of the following equations
\begin{subequations}
\label{eq.g2}
\begin{equation}
\label{eq.ge2}
\bigg[\frac{d}{dr}\mu_t^{-1}\frac{d}{dr}-\frac{l(l+1)}{\mu_pr^2}-\varepsilon_t\zeta^2\bigg]\tilde{e}_{\pm;\zeta,l}(r)=0,
\end{equation}
\begin{equation}
\label{eq.gh2}
\bigg[\frac{d}{dr}\varepsilon_t^{-1}\frac{d}{dr}-\frac{l(l+1)}{\varepsilon_pr^2}-\mu_t\zeta^2\bigg]\tilde{h}_{\pm;\zeta,l}(r)=0.
\end{equation}
\end{subequations}
Suppose the independent solutions for these two equations are, respectively,
\begin{equation}
\label{eq.old10}
\tilde{e}_{\pm;\zeta,l}(r)=re_{\pm;\zeta,l}(r),\ \tilde{h}_{\pm;\zeta,l}(r)=rh_{\pm;\zeta,l}(r),
\end{equation}
with the boundary conditions satisfied, that is, as $r\rightarrow\infty$ ($r\rightarrow0$), $e_+(r)$ and $h_+(r)$ ($e_-(r)$ and $h_-(r)$) approaches finite values (typically zero). Then $g^{\rm E}_{\zeta,l}$ has a simple form
\begin{equation}
g^{\rm E}_{\zeta,l}(r,r')=\frac{\tilde{e}_{+;\zeta,l}(r_>)\tilde{e}_{-;\zeta,l}(r_<)}{W^{\rm E}_{\zeta,l}rr'},
\end{equation}
in which $r_>(r_<)$ is the larger (smaller) one in $r$ and $r'$, and the generalized Wronskian $W^{\rm E}_{\zeta,l}$ is
\begin{equation}
\label{eq.wrons}
W^{\rm E}_{\zeta,l}=\frac{\tilde{e}'_{+;\zeta,l}(r)\tilde{e}_{-;\zeta,l}(r)-\tilde{e}_{+;\zeta,l}(r)\tilde{e}'_{-;\zeta,l}(r)}{\mu_t(\zeta,r)},
\end{equation}
and $g^{\rm H}_{\zeta,l}$ can be obtained by making the substitutions ${\rm E}\rightarrow{\rm H}$, $\varepsilon\leftrightarrow\mu$ and $e\rightarrow h$.

\par With those general arguments above, we focus on the Casimir stress, as well as the corresponding Casimir energy density, of an inhomogeneous dielectric ball. To be more explicit, the TE stress and energy density can be rewritten as (the overlines have been omitted for simplicity without raising any unclarity)
\begin{subequations}
\begin{eqnarray}
&&
T_{E;rr}(\mathbf{r})
=
\sum_{l=1}^{\infty}\frac{\nu}{4\pi r^2}\int\frac{d\zeta}{2\pi}
\frac{\partial\ln[\tilde{e}_{+;\zeta,l}(r_+),\tilde{e}_{-;\zeta,l}(r)]_{\mu_t}}{\partial r_+}
\nonumber\\
&&
\quad
=
\sum_{l=1}^{\infty}\frac{-\nu}{4\pi r^2}\int\frac{d\zeta}{2\pi}
\frac{\partial\ln[\tilde{e}_{+;\zeta,l}(r),\tilde{e}_{-;\zeta,l}(r_-)]_{\mu_t}}{\partial r_-}
,
\end{eqnarray}
\begin{eqnarray}
&&
T_{E;tt}(\mathbf{r})\equiv
T_{E;\theta\theta}(\mathbf{r})=T_{E;\varphi\varphi}(\mathbf{r})=\sum_{l=1}^{\infty}\frac{-\nu}{4\pi r^2}\int\frac{d\zeta}{2\pi}\frac{1}{W^E_{\zeta,l}}
\nonumber\\
&&
\quad\quad
\times\frac{l(l+1)}{\mu_pr^2}\tilde{e}_{+;\zeta,l}(r)\tilde{e}_{-;\zeta,l}(r)
,
\end{eqnarray}
\begin{eqnarray}
&&
U_{E}(\mathbf{r})=\sum_{l=1}^{\infty}\frac{-\nu}{4\pi r^2}\int\frac{d\zeta}{2\pi}\frac{1}{W^E_{\zeta,l}}\frac{\partial}{\partial r}\bigg[\frac{\tilde{e}'_{\pm;\zeta,l}(r)}{\mu_t}\frac{\partial\zeta\tilde{e}_{\mp;\zeta,l}(r)}{\partial\zeta}
\nonumber\\
&&
\quad\quad
-\zeta\tilde{e}_{\pm;\zeta,l}(r)\frac{\partial}{\partial\zeta}\frac{\tilde{e}'_{\mp;\zeta,l}(r)}{\mu_t}\bigg]
,
\end{eqnarray}
where $\nu=l+1/2$, $r_+=r_-=r$ and the bracket $[\cdot,\cdot]_k$ is defined as
\begin{eqnarray}
[f(x),g(y)]_{\mu}=\frac{f'(x)g(y)}{\mu_f(x)}-\frac{f(x)g'(y)}{\mu_g(y)}.
\end{eqnarray}
\end{subequations}
It should be mentioned that the net stresses in the transverse directions here are zero due to the transverse isotropy. The radial pressure within a single body is zero as well. For the interface characterized, for instance, by the discontinuity of electromagnetic properties on its two sides, nontrivial results present.

\par To demonstrate the effects of a spherical interface, we consider a ball (medium 1) in a background (medium 2) with the anisotropy as above. Suppose the radius of the ball is $a$, $\tilde{e}_{\pm}$ in Eq.~\eqref{eq.old10} can be expressed as (the explicit dependence on $\zeta$ and $l$ are implied for the simplicity in symbols when no extra ambiguity is thus introduced)
\begin{subequations}
\label{eq.solutions1}
\begin{equation}
\tilde{e}_{+}(r)=
\left\{
  \begin{array}{cc}
    \tilde{e}_{2+}(r), & r>a, \\
    A^{(e)}_{1+}\tilde{e}_{1+}(r)+B^{(e)}_{1+}\tilde{e}_{1-}(r), & r<a, \\
  \end{array}
\right.
\quad\
\end{equation}
\begin{equation}
\tilde{e}_{-}(r)=
\left\{
  \begin{array}{cc}
    A^{(e)}_{2-}\tilde{e}_{2+}(r)+B^{(e)}_{2-}\tilde{e}_{2-}(r), & r>a, \\
    \tilde{e}_{1-}(r), & r<a, \\
  \end{array}
\right.
\quad\
\end{equation}
\end{subequations}
where the subscript $i=1,2$ signifies the values in the medium $i$. $\tilde{e}_{i\pm}$, satisfying the same boundary conditions as $\tilde{e}_{\pm}$ respectively, are solutions of Eq.~\eqref{eq.ge2} when medium $i$ filling in the whole space. The coefficients now are determined by the continuous conditions across the interface, which lead us to
\begin{subequations}
\begin{equation}
A^{(e)}_{1+}=\frac{[\tilde{e}_{2+}(a),\tilde{e}_{1-}(a)]_{\mu_t}}{W^E_1},\
B^{(e)}_{1+}=\frac{[\tilde{e}_{1+}(a),\tilde{e}_{2+}(a)]_{\mu_t}}{W^E_1},
\end{equation}
\begin{equation}
A^{(e)}_{2-}=\frac{[\tilde{e}_{1-}(a),\tilde{e}_{2-}(a)]_{\mu_t}}{W^E_2},\
B^{(e)}_{2-}=\frac{[\tilde{e}_{2+}(a),\tilde{e}_{1-}(a)]_{\mu_t}}{W^E_2},
\end{equation}
and the Wronskian $W^E$ is
\begin{equation}
W^E=A^{(e)}_{1+}W^E_1=B^{(e)}_{2-}W^E_2.
\end{equation}
\end{subequations}
Focusing on the contributions stemming from the interaction of the ball and the background, that is, ignoring the contributions when each medium filling in the whole space, then for $r<a$, the interaction stress and energy density are
\begin{subequations}
\label{eq.DT0}
\begin{eqnarray}
&&
\Delta T_{E;rr}(\mathbf{r})
=
\sum_{l=1}^{\infty}\frac{-\nu}{4\pi r^2}\int\frac{d\zeta}{2\pi}\frac{R^E_1}{W^E_1}
\bigg\{
\frac{\tilde{e}_{1-}^{\prime2}(r)}{\mu_{1t}}
\nonumber\\
&&
\quad\quad
-
\bigg[
\frac{l(l+1)}{\mu_{1p}r^2}+\varepsilon_{1t}\zeta^2
\bigg]\tilde{e}_{1-}^2(r)
\bigg\}
,
\end{eqnarray}
\begin{eqnarray}
&&
\Delta T_{E;tt}(\mathbf{r})\equiv\Delta T_{E;\theta\theta}(\mathbf{r})=\Delta T_{E;\varphi\varphi}(\mathbf{r})
\nonumber\\
&&
\quad\quad
=\sum_{l=1}^{\infty}\frac{-\nu}{4\pi r^2}\int\frac{d\zeta}{2\pi}\frac{R^E_1}{W^E_1}
\frac{l(l+1)}{\mu_{1p}r^2}\tilde{e}_{1-}^2(r)
,
\end{eqnarray}
\begin{eqnarray}
&&
\Delta U_{E}(\mathbf{r})=\sum_{l=1}^{\infty}\frac{-\nu}{4\pi r^2}\int\frac{d\zeta}{2\pi}\frac{R^E_1\zeta}{W^E_1}\bigg[
\frac{\partial\ln(\mu_{1t}\zeta)}{\partial \zeta}\frac{\partial}{\partial r}\bigg(\frac{\tilde{e}'_{1-}\tilde{e}_{1-}}{\mu_{1t}}\bigg)
\nonumber\\
&&
\quad\quad
-
\frac{\partial(\varepsilon_{1t}\mu_{1t}\zeta^2)}{\partial\zeta}\frac{\tilde{e}_{1-}^2}{\mu_{1t}}
-
\frac{l(l+1)}{\mu_{1p}r^2}\frac{\partial\ln\gamma_{1\mu}^2}{\partial\zeta}\tilde{e}_{1-}^2
\bigg],
\end{eqnarray}
\end{subequations}
while for $r>a$,
\begin{subequations}
\label{eq.DT1}
\begin{eqnarray}
&&
\Delta T_{E;rr}(\mathbf{r})
=
\sum_{l=1}^{\infty}\frac{-\nu}{4\pi r^2}\int\frac{d\zeta}{2\pi}\frac{R^E_2}{W^E_2}
\bigg\{
\frac{\tilde{e}_{2+}^{\prime2}(r)}{\mu_{2t}}
\nonumber\\
&&
\quad\quad
-
\bigg[
\frac{l(l+1)}{\mu_{2p}r^2}+\varepsilon_{2t}\zeta^2
\bigg]\tilde{e}_{2+}^2(r)
\bigg\}
,
\end{eqnarray}
\begin{eqnarray}
&&
\Delta T_{E;tt}(\mathbf{r})\equiv\Delta T_{E;\theta\theta}(\mathbf{r})=\Delta T_{E;\varphi\varphi}(\mathbf{r})
\nonumber\\
&&
\quad\quad
=\sum_{l=1}^{\infty}\frac{-\nu}{4\pi r^2}\int\frac{d\zeta}{2\pi}\frac{R^E_2}{W^E_2}
\frac{l(l+1)}{\mu_{2p}r^2}\tilde{e}_{2+}^2(r)
,
\end{eqnarray}
\begin{eqnarray}
&&
\Delta U_{E}(\mathbf{r})=\sum_{l=1}^{\infty}\frac{-\nu}{4\pi r^2}\int\frac{d\zeta}{2\pi}\frac{R^E_2\zeta}{W^E_2}
\bigg[
\frac{\partial\ln(\mu_{2t}\zeta)}{\partial \zeta}\frac{\partial}{\partial r}\bigg(\frac{\tilde{e}'_{2+}\tilde{e}_{2+}}{\mu_{2t}}\bigg)
\nonumber\\
&&
\quad\quad
-
\frac{\partial(\varepsilon_{2t}\mu_{2t}\zeta^2)}{\partial\zeta}\frac{\tilde{e}_{2+}^2}{\mu_{2t}}
-
\frac{l(l+1)}{\mu_{2p}r^2}\frac{\partial\ln\gamma_{2\mu}^2}{\partial\zeta}\tilde{e}_{2+}^2
\bigg]
,
\end{eqnarray}
\end{subequations}
where $R^E_1=B^{(e)}_{1+}/A^{(e)}_{1+}$, $R^E_2=A^{(e)}_{2-}/B^{(e)}_{2-}$, and the character index $\gamma_{i\mu}^2=\mu_{it}/\mu_{ip}$ explicitly embodies the impact from the anisotropy. Before diving into the exploration on the interaction contributions in the model above, it should be noted that the net TE pressure on the interface at $r=a$ can be written as
\begin{eqnarray}
P_E(a)=\sum_{l=1}^{\infty}\frac{-\nu}{4\pi a^2}\int\frac{d\zeta}{2\pi}\frac{\partial}{\partial a}\ln[\tilde{e}_{2+}(a),\tilde{e}_{1-}(a)]_{\mu_t}
.
\end{eqnarray}
On the other hand, the total TE Casimir energy is derived as
\begin{eqnarray}
\mathcal{U}_{E}=\sum_{l=1}^{\infty}\nu\int\frac{d\zeta}{2\pi}\ln[\tilde{e}_{2+}(a),\tilde{e}_{1-}(a)]_{\mu_t}
,
\end{eqnarray}
implying that the principle of virtual work is satisfied and showing no pressure anomaly. Similar results can be obtained for TM contributions by making the substitutions $\varepsilon\leftrightarrow\mu$, $e\rightarrow h$ and $E\rightarrow H$.

\par In the arguments above, both the inhomogeneity and dispersion are included, introducing plenty of complexities. However, since the impacts from the inhomogeneity, together with those due to the geometry, are to be concentrated on here, we will not take the dispersion into account and work with simpler forms, for instance the energy density can be
\begin{equation}
\Delta U_{E}(\mathbf{r})=\sum_{l=1}^{\infty}\frac{-\nu}{4\pi r^2}\int\frac{d\zeta}{2\pi}\frac{R^E_1}{W^E_1}\bigg[
\frac{\partial}{\partial r}\bigg(\frac{\tilde{e}'_{1-}\tilde{e}_{1-}}{\mu_{1t}}\bigg)
-
2\kappa_1^2\frac{\tilde{e}_{1-}^2}{\mu_{1t}}
\bigg],
\end{equation}
for TE part when $r<a$, and its counterpart when $r>a$
\begin{equation}
\Delta U_{E}(\mathbf{r})=\sum_{l=1}^{\infty}\frac{-\nu}{4\pi r^2}\int\frac{d\zeta}{2\pi}\frac{R^E_2}{W^E_2}
\bigg[
\frac{\partial}{\partial r}\bigg(\frac{\tilde{e}'_{2+}\tilde{e}_{2+}}{\mu_{2t}}\bigg)
-
2\kappa_2^2\frac{\tilde{e}_{2+}^2}{\mu_{2t}}
\bigg]
,
\end{equation}
which also render the energy-momentum tensor of the field traceless. So we mostly consider the stresses in following arguments, and the energy density can be derived straightforwardly.

\par In the next section, we will investigate the behaviors of the stress ad energy density due to the interaction between the ball and the background. Even though Eqs.~\eqref{eq.g2} look simple, the analytical solutions are generally not within the reach. Yet there still exist some analytically solvable models, with which we will start.

\section*{Analyses and Discussions}
\label{sec.ASM}
\par In this section, we chiefly study the interaction Casimir stresses (referred to as Casimir stresses or stresses for short in the following arguments) in three specific models, namely the homogeneous model, $O(r^{-1})$ model and the homogeneous-$O(r^{-2})$ model, together with some relevant cases. Behaviors of Casimir stresses around the surface are emphasized, and the comparison with the planar case is explored. The roles of refractive and anisotropic indices are illustrated as well. General cases are also briefly investigated, in which the origin of the ``softening'' of divergences at the interface by the inhomogeneity is considered.
\subsection{Homogeneous model}
\par As one of the most frequently involved model when studying the Casimir effects in the spherical configuration, we firstly consider a ball in a background, both comprising of homogeneous media, which are isotropic on the transverse directions as depicted in Eq.~\eqref{eq.spepsilon}. $\tilde{e}_{i\pm}$ in Eq.~\eqref{eq.solutions1} can in this case be solved as
\begin{equation}
\label{eq.es}
\tilde{e}_{i+}(r)=e_{\nu_{i\mu}}(\kappa_ir),\ \tilde{e}_{i-}(r)=s_{\nu_{i\mu}}(\kappa_ir),
\end{equation}
where $\kappa_i^2=\varepsilon_{it}\mu_{it}\zeta^2$, $\nu_{i\mu}=\sqrt{\gamma_{i\mu}^2l(l+1)+1/4}$, $\gamma_{i\mu}^2=\mu_{it}/\mu_{ip}$, $e_n(x)$ and $s_n(x)$ are referred to as the modified Ricatti-Bessel functions with the subscript convenient here
\begin{equation}
\label{eq.es}
e_n(x)=\sqrt{\frac{2x}{\pi}}K_{n}(x),\ s_n(x)=\sqrt{\frac{\pi x}{2}}I_{n}(x).
\end{equation}

\par We are thus led to the explicit expressions for the stresses in Eqs.~\eqref{eq.DT0} and \eqref{eq.DT1}. Inside the ball $r<a$, the radial and transverse components of the Casimir stress, the TE contribution as the instance, are accordingly
\begin{subequations}
\label{eq.Tuinside}
\begin{eqnarray}
&&
\Delta T_{{\rm E};rr}
=
\sum_{l=1}^{\infty}\frac{\nu}{4\pi^2n_1a^4}\int_0^{\infty}\frac{dxx}{d^2}
\frac{
[e_{\nu_{1\mu}}(x),e_{\nu_{2\mu}}(n_{21}x)]_{\mu}
}{
[e_{\nu_{2\mu}}(n_{21}x),s_{\nu_{1\mu}}(x)]_{\mu}
}
\nonumber\\
&&
\quad\quad
\times
\bigg[
s_{\nu_{1\mu}}^{\prime2}(xd)
-
s''_{\nu_{1\mu}}(xd)s_{\nu_{1\mu}}(xd)
\bigg]
,\ r<a,
\end{eqnarray}
\begin{eqnarray}
&&
\Delta T_{{\rm E};tt}
=
\sum_{l=1}^{\infty}\frac{\nu}{4\pi^2n_1a^4}\int^{\infty}_0\frac{dxx}{d^2}
\frac{\nu_{1\mu}^2-1/4}{x^2d^2}
s_{\nu_{1\mu}}^2(xd)
\nonumber\\
&&
\quad\quad
\times
\frac{
[e_{\nu_{1\mu}}(x),e_{\nu_{2\mu}}(n_{21}x)]_{\mu}
}{
[e_{\nu_{2\mu}}(n_{21}x),s_{\nu_{1\mu}}(x)]_{\mu}
},\ r<a,
\end{eqnarray}
\end{subequations}
while for those outside the ball,
\begin{subequations}
\label{eq.Tuoutside}
\begin{eqnarray}
&&
\Delta T_{{\rm E};rr}
=
\sum_{l=1}^{\infty}\frac{\nu}{4\pi^2n_2a^4}\int^{\infty}_0\frac{dxx}{d^2}
\frac{
[s_{\nu_{1\mu}}(n_{12}x),s_{\nu_{2\mu}}(x)]_{\mu}
}{
[e_{\nu_{2\mu}}(x),s_{\nu_{1\mu}}(n_{12}x)]_{\mu}
}
\nonumber\\
&&
\quad\quad
\times
\bigg[
e_{\nu_{2\mu}}^{\prime2}(xd)
-
e''_{\nu_{2\mu}}(xd)e_{\nu_{2\mu}}(xd)
\bigg]
,\ r>a,
\end{eqnarray}
\begin{eqnarray}
&&
\Delta T_{{\rm E};tt}
=
\sum_{l=1}^{\infty}\frac{\nu}{4\pi^2n_2a^4}\int^{\infty}_0\frac{dxx}{d^2}\frac{\nu_{2,\mu}^2-1/4}{x^2d^2}
e_{\nu_{2\mu}}^2(xd)
\nonumber\\
&&
\quad\quad
\times
\frac{
[s_{\nu_{1\mu}}(n_{12}x),s_{\nu_{2\mu}}(x)]_{\mu}
}{
[e_{\nu_{2\mu}}(x),s_{\nu_{1\mu}}(n_{12}x)]_{\mu}
},\ r>a,
\end{eqnarray}
\end{subequations}
where $n_i=\sqrt{\varepsilon_i\mu_i},\ n_{ij}=n_i/n_j$, and $d=r/a$. Corresponding TM contributions can be obtained with the substitutions $\rm E\rightarrow H$ and $\varepsilon\leftrightarrow\mu$. Evidently Eqs.~\eqref{eq.Tuinside} and \eqref{eq.Tuoutside} are too complicated to be solved analytically in general, and only elaborately designed numerical evaluations can give us an overview on the general behaviors of these quantities.

\par Nevertheless, analytical considerations in some limiting cases could still unveil some helpful physics hidden in expressions above. For instance, in the planar limit $a\rightarrow\infty$, the curvature of the surface goes to zero. We employ the uniform asymptotic expansions (UAE) to capture the main features of the modified Ricatti-Bessel functions as did in Refs.~\cite{milton1980semiclassical,milton2018casimir}. Then inside the ball $r<a$, the TE radial and transverse stresses for the limit $a\rightarrow\infty,\ d\rightarrow1$ with $a-r$ fixed, to the leading order, are
\begin{subequations}
\label{eq.planar}
\begin{eqnarray}
&&
\Delta T_{{\rm E};rr}
\sim
\frac{-1}{a^2(r-a)^2}
\int_{0}^{\infty}\frac{dkk}{4\pi^2n_1}\int^{\infty}_0\frac{dx}{d^3}
\frac{e^{-2\sqrt{\gamma_{1\mu}^2k^2+x^2}}}{8\sqrt{\gamma_{1\mu}^2k^2+x^2d^2}}
\nonumber\\
&&\quad
\times\frac{
y_{\mu}\sqrt{\gamma_{2\mu}^2k^2+n_{21}^2x^2}
-
\sqrt{\gamma_{1\mu}^2k^2+x^2}
}{ 
y_{\mu}\sqrt{\gamma_{2\mu}^2k^2+n_{21}^2x^2}
+
\sqrt{\gamma_{1\mu}^2k^2+x^2}
},\ y_\mu=\frac{\mu_{1t}}{\mu_{2t}},\quad\quad\quad
\end{eqnarray}
\begin{eqnarray}
&&
\Delta T_{{\rm E};tt}
\sim
\frac{-1}{(r-a)^4}\int_{0}^{\infty}\frac{dkk}{4\pi^2n_1}\int^{\infty}_0\frac{dx}{d^3}
\frac{\gamma_{1\mu}^2k^2e^{-2\sqrt{\gamma_{1\mu}^2k^2+x^2}}}{2\sqrt{\gamma_{1\mu}^2k^2+x^2d^2}}
\nonumber\\
&&\quad\quad
\times
\frac{
y_{\mu}\sqrt{\gamma_{2\mu}^2k^2+n_{21}^2x^2}
-
\sqrt{\gamma_{1\mu}^2k^2+x^2}
}{ 
y_{\mu}\sqrt{\gamma_{2\mu}^2k^2+n_{21}^2x^2}
+
\sqrt{\gamma_{1\mu}^2k^2+x^2}
}.
\end{eqnarray}
\end{subequations}
in which the substitutions $\sum\limits_{l=1}\nu/a^2\rightarrow\int\limits^{\infty}_0dkk,\ \nu/a\rightarrow k$ is utilized. We immediately see that although the difference on the speed of light denoted by $n_{21}$, and the anisotropy denoted by $\gamma_{i\mu}$ do introduce notable extra complexities, they hardly cause any divergences at the surface. It is still the vanishing distance from the surface that gives rise to the surface divergence. Another observation is that the spherical geometry of the whole system significantly impacts the Casimir interaction. In fact, results in Eq.~\eqref{eq.planar} are exactly consistent with their corresponding planar counterparts, as shown in the Appendix~\ref{ApA}. This is achieved only owing to the zero-curvature limit, where the interaction between the ball and the background approaches that in the planar case. Generally the influences from the curvature, signified by terms depending on the radius $a$ here, are vital, as we shall see in the behaviors of stresses close to the surface as below.

\par First, we consider configurations involving the perfect conductor (with the permittivity approaching infinity and boundary conditions of the ideal conductor applied), that is, the ball in the perfectly conducting background (case I) and the perfectly conducting ball in the homogeneous background (case II). For case I, the Casimir stresses within the ball close to the surface behave as ($q=(r-a)/a$)
\begin{subequations}
\label{eq.TpI}
\begin{eqnarray}
&&
\Delta T_{{\rm E};rr}
\sim
\sum_{l=1}^{\infty}\frac{-\nu}{8\pi^2n_1a^4d^3}\int_0^{\infty}dx
e^{-2\nu_{1\mu}|q|/p_1}
\bigg(
p_1^2
\nonumber\\
&&\quad
+\frac{3}{2\nu_{1\mu}}p_1^3-\frac{p_1^5}{\nu_{1\mu}}-2p_1^2q+2p_1^4q
\bigg)
,\ p_1=\frac{\nu_{1\mu}}{\sqrt{\nu_{1\mu}^2+x^2}},
\nonumber\\
&=&
\frac{-1}{8\pi^2n_1\gamma_{1\mu}^2a^4d^3|q|^3}
\bigg\{
\frac{1}{6}+\frac{3}{20}|q|
\bigg\}
,\ r<a,
\end{eqnarray}
\begin{eqnarray}
&&
\Delta T_{{\rm E};tt}
\sim
\sum_{l=1}^{\infty}\frac{-\nu}{8\pi^2n_1a^4d^3}\int^{\infty}_0dx
e^{-2\nu_{1\mu}|q|/p_1}
\bigg(
\nu_{1\mu}p_1
\\
&&\quad
-p_1\nu_{1\mu}q
+p_1^3\nu_{1\mu}q
\bigg)
=
\frac{-1}{8\pi^2n_1\gamma_{1\mu}^2a^4d^3|q|^3}
\bigg(
\frac{1}{4|q|}+\frac{1}{20}
\bigg)
,\nonumber
\end{eqnarray}
\begin{eqnarray}
&&
\Delta T_{{\rm H};rr}
\sim
\sum_{l=1}^{\infty}\frac{-\nu}{8\pi^2n_1a^4d^3}\int_0^{\infty}dx
e^{-2\nu_{1\varepsilon}|q|/p_2}
\bigg(
-p_2^2
\nonumber\\
&&\quad
-\frac{3}{2}\frac{p_2^3}{\nu_{1\varepsilon}}+2\frac{p_2^5}{\nu_{1\varepsilon}}
+2p_2^2q-2p_2^4q
\bigg),\ p_2=\frac{\nu_{1\varepsilon}}{\sqrt{\nu_{1\varepsilon}^2+x^2}},
\nonumber\\
&=&
\frac{-1}{8\pi^2n_1\gamma_{1\varepsilon}^2a^4d^3|q|^3}
\bigg\{
-\frac{1}{6}-\frac{1}{20}|q|
\bigg\}
,
\end{eqnarray}
\begin{eqnarray}
&&
\Delta T_{{\rm H};tt}
\sim
\sum_{l=1}^{\infty}\frac{-\nu}{8\pi^2n_1a^4}\int^{\infty}_0\frac{dx}{d^3}
e^{-2\nu_{1\varepsilon}|q|/p_2}
\bigg(-\nu_{1\varepsilon}p_2
\nonumber\\
&&\quad
+p_2^4+q\nu_{1\varepsilon}p_2-q\nu_{1\varepsilon}p_2^3\bigg)
\nonumber\\
&=&
\frac{-1}{8\pi^2n_1\gamma_{1\varepsilon}^2a^4d^3|q|^3}
\bigg(-\frac{1}{4|q|}+\frac{13}{60}\bigg)
,\ r<a.
\end{eqnarray}
\end{subequations}
For case II, the Casimir stresses on the outer side of the interface behave as
\begin{subequations}
\label{eq.TpII}
\begin{eqnarray}
\label{eq.caseIITErr}
&&
\Delta T_{{\rm E};rr}
\sim
\sum_{l=1}^{\infty}\frac{-\nu}{8\pi^2n_2a^4d^3}\int^{\infty}_0dx
e^{-2\nu_{2\mu}q/p_3}
\bigg(
-p_3^2
\nonumber\\
&&\quad
+2qp_3^2-2qp_3^4+\frac{3}{2}\frac{p_3^3}{\nu_{2\mu}}-\frac{p_3^5}{\nu_{2\mu}}
\bigg),\ p_3=\frac{\nu_{2\mu}}{\sqrt{\nu_{2\mu}^2+x^2}},
\nonumber\\
&=&
\frac{-1}{8\pi^2n_2\gamma_{2\mu}^2a^4d^3q^3}
\bigg(
-\frac{1}{6}+q\frac{11}{60}
\bigg)
,
\end{eqnarray}
\begin{eqnarray}
&&
\Delta T_{{\rm E};tt}
\sim
\sum_{l=1}^{\infty}\frac{-\nu}{8\pi^2n_2a^4}\int^{\infty}_0\frac{dx}{d^3}
e^{-\frac{2\nu_{2\mu}q}{p_3}}
\bigg(\nu_{2\mu}p_3-q\nu_{2\mu}p_3
\nonumber\\
&&\quad
+q\nu_{2\mu}p_3^3\bigg)
=
\frac{-1}{8\pi^2n_2\gamma_{2\mu}^2a^4d^3q^3}
\bigg(\frac{1}{4q}-\frac{1}{20}\bigg)
,
\end{eqnarray}
\begin{eqnarray}
\label{eq.caseIITMrr}
&&
\Delta T_{{\rm H};rr}
\sim
\sum_{l=1}^{\infty}\frac{-\nu}{8\pi^2n_2a^4d^3}\int^{\infty}_0dx
e^{-2\nu_{2\varepsilon}q/p_4}
\bigg(p_4^2-2qp_4^2
\nonumber\\
&&\quad
+2qp_4^4-\frac{3}{2}\frac{p_4^3}{\nu_{2\varepsilon}}+2\frac{p_4^5}{\nu_{2\varepsilon}}\bigg)
,\ p_4=\frac{\nu_{2\varepsilon}}{\sqrt{\nu_{2\varepsilon}^2+x^2}},
\nonumber\\
&=&
\frac{-1}{8\pi^2n_2\gamma_{2\varepsilon}^2a^4d^3q^3}
\bigg(\frac{1}{6}-\frac{1}{20}q\bigg)
,
\end{eqnarray}
\begin{eqnarray}
&&
\Delta T_{{\rm H};tt}
\sim
\sum_{l=1}^{\infty}\frac{-\nu}{8\pi^2n_2a^4d^3}\int^{\infty}_0dx
e^{-\frac{2\nu_{2\varepsilon}q}{p_4}}
\bigg(q\nu_{2\varepsilon}p_4-\nu_{2\varepsilon}p_4
\nonumber\\
&&\
-p_4^4-q\nu_{2\varepsilon}p_4^3\bigg)
=
\frac{-1}{8\pi^2n_2\gamma_{2\varepsilon}^2a^4d^3q^3}
\bigg(-\frac{1}{4q}-\frac{1}{12}\bigg)
.
\end{eqnarray}
\end{subequations}
The leading terms of stresses consist exactly with the corresponding planar cases as the perfectly conducting limit applied to Eq.~\eqref{eq.pltr} in Appendix \ref{ApA}.
Notably, the TE and TM contributions to these cancel, softening the divergences as $q\rightarrow0$, and the transverse stresses divergences faster than their radial counterparts. Yet it is the radial stresses, reflected as the net force at the surface, that might exert observable effects. As an instance, Milton~\cite{milton1980semiclassical} considered the net force on perfectly conducting ball. If we use the temporal point-splitting $\tau$ exactly at the surface as in Ref.~\cite{milton1980semiclassical}, then we should set $q=0$ and combine Eqs.~\eqref{eq.caseIITErr} and \eqref{eq.caseIITMrr}, which give us the force density at the surface $\mathcal{F}=-\Delta T_{{\rm E};rr}-\Delta T_{{\rm H};rr}$
\begin{eqnarray}
\label{eq.FFF}
&&
\mathcal{F}=
\sum_{l=1}^{\infty}\frac{\nu}{8\pi^2n_2a^4}\int^{\infty}_0dz
\cos(\nu z\tau)
\frac{1}{(1+z^2)^{\frac{5}{2}}}
\nonumber\\
&\sim&
\int_{0}^{\infty}\frac{dkk^3K_2(k)}{24\pi^2n_2a^4\tau^2}
=
\frac{1}{3\pi^2n_2a^4\tau^2}
\nonumber\\
&\sim&
\frac{1}{3\pi^2n_2a^4\tau^2}-\frac{1}{32\pi^2n_2a^4}
,
\end{eqnarray}
where the background is assumed isotropic. The second line of Eq.~\eqref{eq.FFF} is obtained with the same substitution in Eq.~\eqref{eq.planar} except for changing $a^{-1}$ to $\tau$, while the third line is derived with the Euler-Maclaurin (EM) formula. Both methods give the same leading behavior, i.e. the divergent term as Eq.~(53) of Ref.~\cite{milton1980semiclassical}, and the EM formula also captures the finite term. On the other hand, if we take $q\neq0,\tau=0$, then force density on the surface is
\begin{eqnarray}
\label{eq.FFF2}
\mathcal{F}
=
\frac{1}{15\pi^2n_2a^4q^2}\bigg|_{q\rightarrow0}
.
\end{eqnarray}
The leading terms in Eqs.~\eqref{eq.FFF} and \eqref{eq.FFF2} diverge both in the inverse second-order of their ``regulators'', i.e. $\tau$ and $q$ respectively, and the impact from the spherical geometry is clearly shown in the factor $a^{-4}$. This might imply the local interaction is the main contributor, yet the spherical geometry is inevitable as the parameter $q$ is the distance to the surface scaled by its radius $a$, besides the explicit $a^4$ in the denominator. Differences on factors between these two leading terms suggest the importance of the details of the surface in possible practical applications.
\begin{figure*}
    \centering
    \includegraphics[width=2.0\columnwidth]{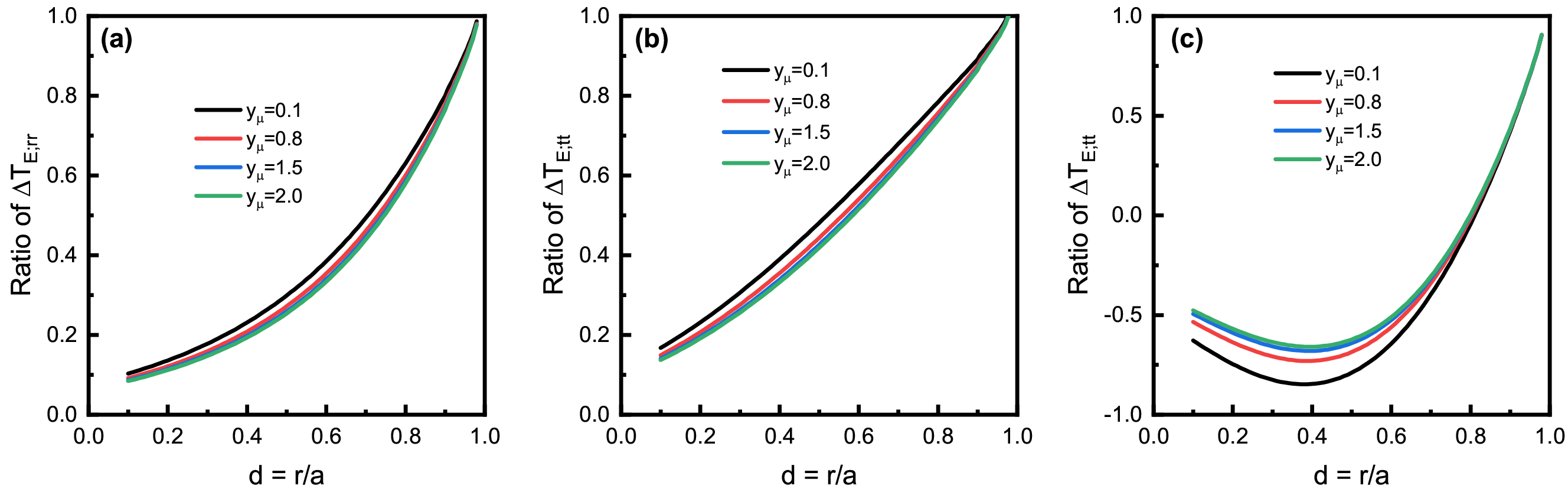}
    \caption{The ratios of Casimir stresses in Eq.~\eqref{eq.vicinity2} over their counterparts in Eq.~\eqref{eq.Tuinside} for different $y_\mu$ plotted as functions of $d=r/a$ with $n_{21}=1$ and $\gamma_{1\mu}=\gamma_{2\mu}=1$. In \textbf{b}, only the first term in Eq.~\eqref{eq.vicinity2.b} is counted, while in \textbf{c} both terms there are included.}
    \label{fig:ratioTEinside}
\end{figure*}

\par Next, we consider the more general case, that is, the homogeneous configuration above. In the vicinity of the surface $r=a$, Eq.~\eqref{eq.ge2} for medium $i$ can be approximated, with $q$ kept to the second order in the equation, as
\begin{eqnarray}
\label{eq.equation02}
\bigg[\frac{d^2}{dq^2}-\xi_{i\mu}^2+(\nu_{i\mu}^2-1/4)(2q-3q^2)\bigg]\tilde{e}_{i\pm}(r)\approx0,
\end{eqnarray}
where $\xi_{i\mu}=\sqrt{\nu_{i\mu}^2-1/4+\varepsilon_{it}\mu_{it}\zeta^2a^2}$ and $q=(r-a)/a\rightarrow0$. Write $\tilde{e}_{i\pm}(r)=\tilde{e}_{i\pm}(q)$ as
\begin{eqnarray}
\tilde{e}_{i\pm}(q)=e^{\mp\xi_{i\mu}q}\sum_{n=0}^{\infty}b_{n\pm}q^n,\ b_{0\pm}=1,
\end{eqnarray}
which, up to $O(q^4)$, can be solved as
\begin{eqnarray}
\label{eq.TEvicinity}
&&
\tilde{e}_{i\pm}(q)
=
e^{\mp\xi_{i\mu}q}
\bigg[
1+\frac{\nu_{i\mu}^2-1/4}{2\xi_{i\mu}^2}q\pm\frac{\nu_{i\mu}^2-1/4}{2\xi_{i\mu}}q^2
\nonumber\\
&&\quad\quad
+\frac{\nu_{i\mu}^2-1/4}{12}\bigg(3-\frac{\nu_{i\mu}^2-1/4}{\xi_{i\mu}^2}\bigg)q^4
\bigg],
\end{eqnarray}
with the Wronskian being
\begin{eqnarray}
&&
W^E_i=-\frac{2\xi_{i\mu}}{\mu_{it}a}\bigg\{1
-\frac{(\nu_{i\mu}^2-1/4)[5(\nu_{i\mu}^2-1/4)-6\xi_{i\mu}^2]}{12\xi_{i\mu}^2}q^4
\nonumber\\
&&\quad\quad
-o(q^4)\bigg\}\approx-\frac{2\xi_{i\mu}}{\mu_{it}a}.
\end{eqnarray}
The most divergent terms of TE Casimir stresses inside the ball when $q\rightarrow0$ are
\begin{subequations}
\label{eq.vicinity}
\begin{eqnarray}
\label{eq.vicinity.a}
&&
\Delta T_{E;rr}
\sim
\frac{-1}{4\pi a^4|q|^3}\int_{0}^{\infty}dkk\int^{\infty}_0\frac{d\zeta}{2\pi}\frac{\gamma_{1\mu}^2k^2}{\kappa_{1\mu}^2}
\nonumber\\
&&\quad\quad
\times
\frac{y_\mu\kappa_{2\mu}-\kappa_{1\mu}}{y_\mu\kappa_{2\mu}+\kappa_{1\mu}}e^{-2\kappa_{1\mu}}
,
\end{eqnarray}
\begin{eqnarray}
\label{eq.vicinity.b}
&&
\Delta T_{E;tt}
\sim
\int_{0}^{\infty}\frac{-dkk}{4\pi a^4q^4}\int_{0}^{\infty}\frac{d\zeta}{2\pi}
\frac{\gamma_{1\mu}^2k^2}{\kappa_{1\mu}}
\frac{y_\mu\kappa_{2\mu}-\kappa_{1\mu}}{y_\mu\kappa_{2\mu}+\kappa_{1\mu}}
e^{-2\kappa_{1\mu}}
\nonumber\\
&&\quad
+
\int_{0}^{\infty}\frac{-dkk}{4\pi a^4|q|^3}\int_{0}^{\infty}\frac{d\zeta}{2\pi}
\bigg[
\frac{y_\mu\kappa_{2\mu}-\kappa_{1\mu}}{y_\mu\kappa_{2\mu}+\kappa_{1\mu}}
\nonumber\\
&&\quad
\times
\frac{4\gamma_{1\mu}^2k^2\kappa_{1\mu}^2-\gamma_{1\mu}^4k^4+\gamma_{1\mu}^4k^4\kappa_{1\mu}}{-\kappa_{1\mu}^3}
\nonumber\\
&&\quad
+
\frac{\gamma_{1\mu}^4k^4\kappa_{2\mu}^2-\gamma_{1\mu}^2\gamma_{2\mu}^2k^4y_\mu\kappa_{1\mu}^2}{\kappa_{1\mu}^2\kappa_{2\mu}^2(y_\mu\kappa_{2\mu}+\kappa_{1\mu})^2}
\bigg]e^{-2\kappa_{1\mu}}
,
\end{eqnarray}
\end{subequations}
where $\kappa_{i\mu}=\sqrt{\gamma_{i\mu}^2k^2+n_i^2\zeta^2}$. As the Eq.~\eqref{eq.TpI} and \eqref{eq.TpII}, the leading contributions, that is, $0$ for $\Delta T_{E;rr}$ and the first term of Eq.~\eqref{eq.vicinity.b} for $\Delta T_{E;tt}$, are exactly those due to the interaction in the planar case and the influences due to the spherical geometry are completely absent. This implies contributions from the local interaction between the ball and background is dominating near the surface, since only the interaction within the small region across the surface can be scarcely affected by the geometry of the system. Influences of the spherical geometry, signified by the explicit dependence on the $O(a^{-1})$ and diverging as $O(q^{-3})$ with $q\rightarrow0$, appear as the secondary terms. As can be seen clearly, in the close vicinity of the surface, we can rarely get rid of impacts due to the geometric configuration of the whole system, especially for the radial part of the stress tensor, which otherwise should be zero as in the planar case. Since the refractive and anisotropic index of the medium here only scale the wavenumber $k$ the frequency $\zeta$, to catch a glimpse on the behaviors of stresses in Eq.~\eqref{eq.vicinity}, we can take $\gamma_{1\mu}=\gamma_{2\mu}=\gamma,\ n_1=n_2=n$, and $y_\mu\neq1$, then
\begin{subequations}
\label{eq.vicinity2}
\begin{eqnarray}
\label{eq.vicinity2.a}
&&
\Delta T_{E;rr}
\sim
\frac{-1}{48\pi^2\gamma^2n a^4|q|^3}
\frac{y_\mu-1}{y_\mu+1}
,
\end{eqnarray}
\begin{eqnarray}
\label{eq.vicinity2.b}
&&
\Delta T_{E;tt}
\sim
\frac{-1}{32\pi^2\gamma^2na^4q^4}\frac{y_\mu-1}{y_\mu+1}
\nonumber\\
&&\quad
+
\frac{(9y_\mu+10)(y_{\mu}-1)}{60\pi^2\gamma^2na^4|q|^3(y_\mu+1)^2}
.
\end{eqnarray}
\end{subequations}
As shown by Fig.~\ref{fig:ratioTEinside}, only extremely close to the surface, Eq.~\eqref{eq.vicinity2} accurately depicts the behaviors of Casimir stresses, which are fully expressed by Eq.~\eqref{eq.Tuinside}. At the point relatively far away from the surface, the deviation from the results in Eq.~\eqref{eq.Tuinside} would be evident, indicating the highly nontrivial contribution from the spherical geometry. This is even more evidently demonstrated in Fig.~\ref{fig:ratioTEinside}c, where the second term of Eq.~\eqref{eq.vicinity2.b} gives raise to the non-monotonicity in the curve. The TM contributions and Casimir stresses outside the ball present similar results, so we do not give an detailed discussion for brevity, and the following exhibition is enough to show a picture.

\par
\begin{figure*}
    \centering
    \includegraphics[width=2\columnwidth]{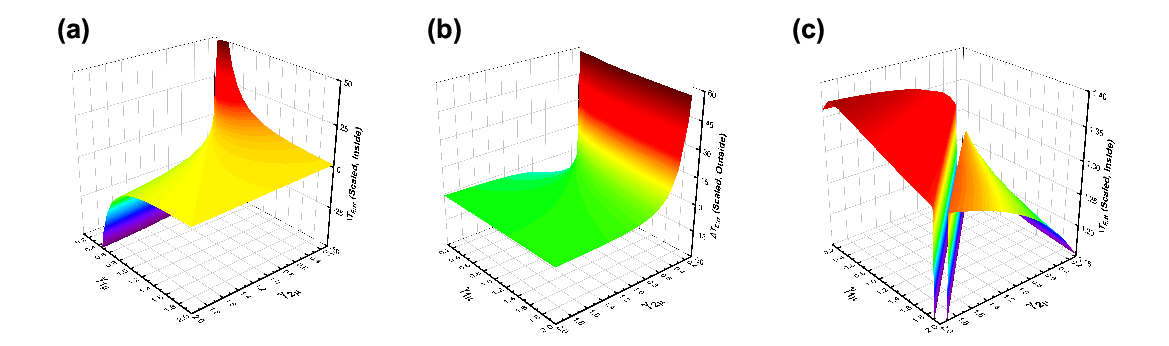}
    \caption{The TE Casimir stresses plotted as functions of the anisotropic indices $\gamma_{1\mu}$ and $\gamma_{2\mu}$ with $n_{21}=1$, $y_{\mu}=1$ and $d=0.8$ (inside, \textbf{a}) or $1.2$ (outside, \textbf{b} and \textbf{c}). The radial Casimir stresses inside and outside the ball are scaled by $1/4\pi^2n_1a^4$ and $1/4\pi^2n_2a^4$, respectively, while the transverse Casimir stresses are scaled by their corresponding planar counterparts.}
    \label{fig:schematic}
\end{figure*}

 As stated above, the numerical evaluations should be resorted to, in the aim of catching the general behaviors of the Casimir stresses. Notably, parameters, such as the anisotropic indices, form a considerably large space, and introduce plenty of diversities to the general behaviors of Casimir stresses. To demonstrate these highly nontrivial impacts succinctly, in Fig.~\ref{fig:schematic} we show the TE Casimir stresses as functions of anisotropic indices for example. Fig.~\ref{fig:schematic}a and Fig.~\ref{fig:schematic}b show that although both anisotropic indices of two touching media act as the scaling factors for the angular index, apparently distinct dependence on anisotropic indices for $\Delta T_{E;rr}$ inside and outside the ball (Fig.~\ref{fig:schematic}a and \ref{fig:schematic}b) occurs. Moreover, for the transverse stress $\Delta T_{E,tt}$, the non-monotonicity in Fig.~\ref{fig:schematic}c around $\gamma_{1\mu}=0.2$ is seen. Even with the diversity introduced by parameters like the anisotropic indices, as well as the long-known refractive indices~\cite{milton2018casimir}, these parameters still do not directly lead to divergences as can be further illustrated via the analytic expressions and further numerics. Yet they can significantly change the structure of divergences, which implies that properties of the media, combined with the geometric configuration, should be responsible for the complications in properly interpreting the Casimir interaction and the relevant physical quantities in the spherical systems. Among various properties of the media, the inhomogeneity is one of the most primary. In the following, we first discuss the role of inhomogeneity by introducing two analytically solvable inhomogeneous models, then study the general behaviors.

\subsection{Exact solvable inhomogeneous models}
\par As can be seen above, such as in Eqs.~\eqref{eq.spGamma1}, \eqref{eq.spGamma2}, \eqref{eq.g2} etc., for a system with the spherical symmetry preserved, the spherical wave modes, labeled by the angular index, can be employed to construct the Green's dyadic. Actually no matter we express the Green's dyadic with the spherical wave modes, plane wave modes or other possible mode structures, the same results should be derived for the same system, as has been shown in Appendix~\ref{ApB} for the bulk Casimir stresses in an isotropic homogeneous background. This spherical mode structure used in this work is just for simplicity and clarity. It can be fairly anticipated that for a given mode structure, the form of each mode included, which essentially depends on the inhomogeneity of the media, will substantially change the Casimir interactions within a single medium (like demonstrated in Appendix~\ref{ApB}) and between two media.

\par The wave form of a spherical mode is not necessarily simple, even for a medium with the simple permittivity and permeability like the homogeneous case above. To catch a glimpse on the influences due to the inhomogeneity, yet keep expressions concise, we consider the model where the ball and background comprise of media with the permittivity and permeability of the form
\begin{eqnarray}
\label{eq.md1}
\varepsilon_{it}(\mu_{it})=\frac{\varepsilon_{it0}(\mu_{it0})}{r}a,\ \varepsilon_{ip}(\mu_{ip})=\frac{\varepsilon_{ip0}(\mu_{ip0})}{r}a,
\end{eqnarray}
which are expected to ``compensate'' the effects of the geometry to some extent, and bring about the wave form closer to plane wave. $a$ acts as the factor rendering $\varepsilon_{i0}$ and $\mu_{i0}$ dimensionless. Then $\tilde{e}_{i\pm}$ in Eq.~\eqref{eq.solutions1} can be solved as
\begin{eqnarray}
\tilde{e}_{i\pm;\zeta,l}(r)=r^{\mp\xi_{i\mu}},\ \xi_{i\mu}=\sqrt{\nu_{i\mu}^2-1/4+\varepsilon_{it0}\mu_{it0}\zeta^2a^2},\quad\quad
\end{eqnarray}
where $\nu_{i\mu}=\sqrt{\gamma_{i\mu}^2l(l+1)+1/4}$ and $\gamma_{i\mu}^2=\mu_{it0}/\mu_{ip0}$. Then the TE Casimir stresses inside the ball are $\Delta T_{E;rr}=0$, exactly the same as the planar case, and
\begin{eqnarray}
\label{eq.ana1}
&&
\Delta T_{E;tt}
=
\sum_{l=1}^{\infty}\frac{\nu}{4\pi r^3}\int^\infty_0\frac{d\zeta}{2\pi}
\frac{\xi_{1\mu}-y_{\mu}\xi_{2\mu}}{y_{\mu}\xi_{2\mu}+\xi_{1\mu}}
\bigg(\frac{r}{a}\bigg)^{2\xi_{1\mu}}
\nonumber\\
&&\quad\quad
\times
\frac{\nu_{1\mu}^2-1/4}{\xi_{1\mu}}
,\ y_\mu=\frac{\mu_{1t}}{\mu_{2t}}=\frac{\mu_{1t0}}{\mu_{2t0}}.
\end{eqnarray}
With this kind of inhomogeneous media, the stresses are more in line with their planar counterparts than those for the homogeneous case in Eq.~\eqref{eq.Tuinside}. In the vicinity of the surface, $\Delta T_{E;tt}$ behaves as
\begin{eqnarray}
&&
\Delta T_{E;tt}
=
\int_{0}^{\infty}\frac{dkk}{4\pi a^4}\int^\infty_0\frac{d\zeta}{2\pi}
\frac{\kappa_{1\mu}-y_{\mu}\kappa_{2\mu}}{y_{\mu}\kappa_{2\mu}+\kappa_{1\mu}}
e^{-2\kappa_{1\mu}}
\nonumber\\
&&\quad\quad
\times
\frac{\gamma_{1\mu}^2k^2}{\kappa_{1\mu}}(q^{-4}-3q^{-3}).
\end{eqnarray}
The TM contributions can be obtained by making the substitutions $E\rightarrow H,\ \varepsilon\leftrightarrow\mu$, and it is easy to check that the Casimir stresses outside the ball give similar results.

\par The observation that the dividing the spatial coordinates as two transverse and one radial suggests another analogy to the planar case. Suppose the media have the permittivity and permeability
\begin{eqnarray}
\label{eq.md2}
\varepsilon_t(\mu_t)=\varepsilon_{t0}(\mu_{t0}),\ \varepsilon_p(\mu_p)=\frac{\varepsilon_{p0}(\mu_{p0})}{r^2}a^2,
\end{eqnarray}
in which the transverse components are homogeneous, while the radial parts are of $r^{-2}$ order inhomogeneity, exactly canceling the intrinsic spatial varying in the spherical wave equation. The solutions of Eq.~\eqref{eq.solutions1} are then
\begin{eqnarray}
\tilde{e}_{\pm;\zeta,l}(r)=e^{\mp\xi_{\mu}r},\ \xi_{\mu}=\sqrt{\frac{\nu_{\mu}-1/4}{a^2}+\varepsilon_{t0}\mu_{t0}\zeta^2},
\end{eqnarray}
leading us to $\Delta T_{E;rr}=0$ as expected, and $\Delta T_{E;tt}$ is
\begin{eqnarray}
\label{eq.ana2}
&&
\Delta T_{E;tt}
=
\sum_{l=1}^{\infty}\frac{\nu}{4\pi r^2}\int^\infty_0\frac{d\zeta}{2\pi}
\frac{\xi_{1\mu}-y_{\mu}\xi_{2\mu}}{y_{\mu}\xi_{2\mu}+\xi_{1\mu}}
e^{-2\xi_{1\mu}(a-r)}
\nonumber\\
&&\quad\quad
\times
\frac{\nu_{1\mu}^2-1/4}{\xi_{1\mu}a^2}
,\ y_\mu=\frac{\mu_{1t}}{\mu_{2t}}=\frac{\mu_{1t0}}{\mu_{2t0}}.
\end{eqnarray}
The leading divergent term in both these inhomogeneous models in the vicinity of the surface are just the same as the corresponding homogeneous planar case, and depend only on values of permitivities and permeabilities at the surface, justifying the statement that the leading divergence are caused by the local interaction at the surface.

\par
\begin{figure}
    \centering
    \includegraphics[width=1.0\columnwidth]{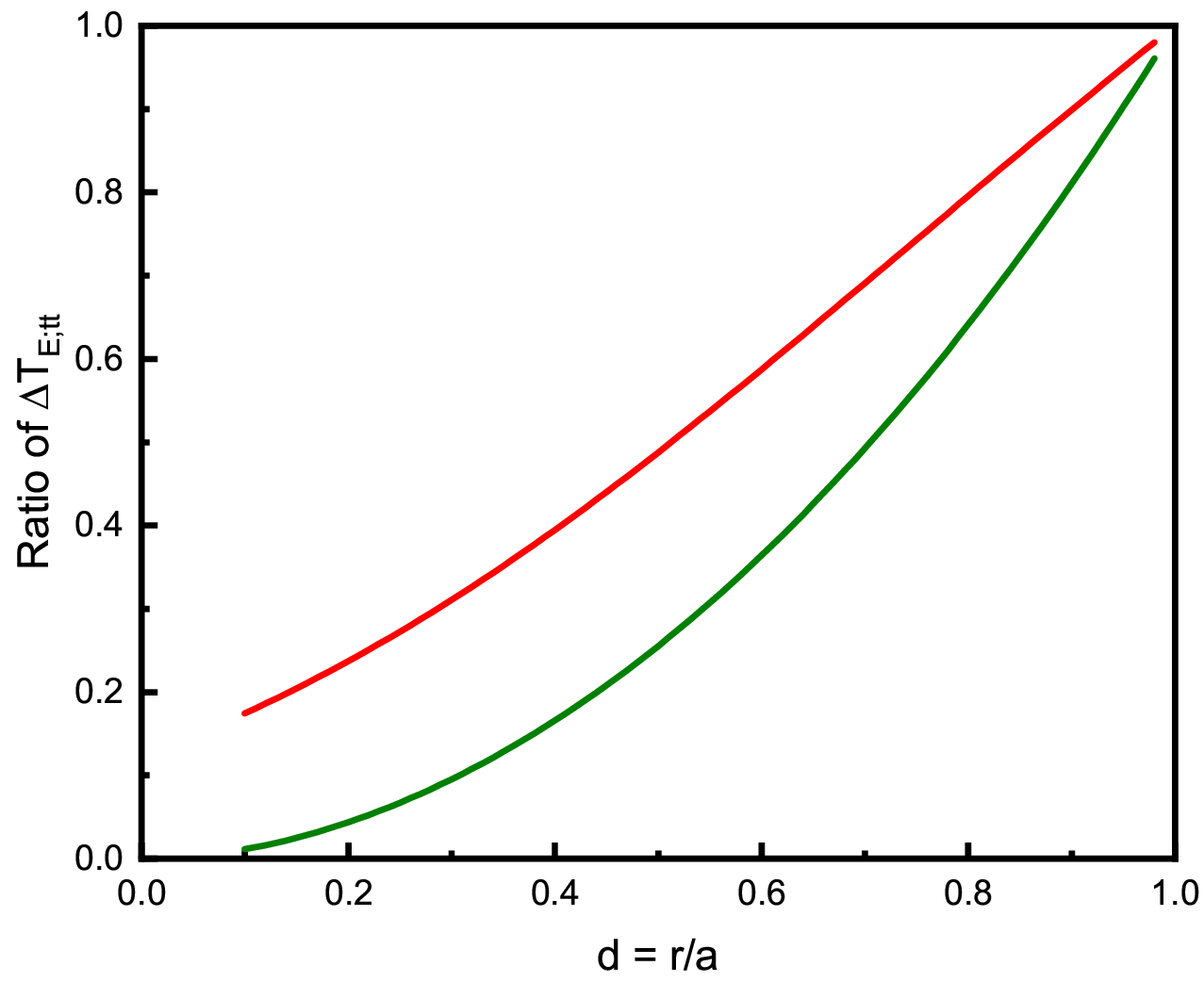}
    \caption{The ratios of planar Casimir stress, i.e. the the first term in Eq.~\eqref{eq.vicinity2.b}, over Casimir stresses in Eq.~\eqref{eq.ana1} (red) and Eq.~\eqref{eq.ana2} (green) plotted as functions of $d=r/a$ with $n_{21}=1$ and $\gamma_{1\mu}=\gamma_{2\mu}=1$. In this case, the ratios here are independent of $y_{\mu}$.}
    \label{fig:A3}
\end{figure}

 As shown above, the impacts on Casimir stresses in the spherical configuration can largely owe to the joint influence from the inhomogeneity of the media and the spherical geometry, and the reduction or cancelation on the intrinsic $O(r^{-2})$ spatial varying in the spherical wave equation from the inhomogeneity of the media can, to a considerable extent, make the wave form of the spherical mode similar to that in the planar configuration. However, this similarity is superficial, since difference in the mode structure is also vital, which is schematically displayed in Fig.~\ref{fig:A3}. There we eliminate the scaling effect of the refractive and anisotropic indices as in Fig.~\ref{fig:ratioTEinside}, and find the convergence to the planar results only occurs at the surface. The discrepancy is evident in Fig.~\ref{fig:A3}. Moreover, it is the model in Eq.~\eqref{eq.md1}, rather than Eq.~\eqref{eq.md2}, that is ``closer'' to the planar case. Interesting phenomena in practical scenarios, such as the designed repulsion and attraction, can be promising by tuning the inhomogeneity of the media. But whatever realistic applications might be in the future, the divergent behaviors of Casimir stresses near the surface should be figured out and interpreted properly. In this aim, we briefly evaluate the general behaviors near the surface of the ball next.

\subsection*{General behaviors near the surface}
\par To make sense of the Casimir interaction and derive relevant measurable physical quantities, divergences have been well recognized as what should be properly dealt with or even systematically renormalized. Divergences are of multiple types, and in this work only two are involved, namely the bulk divergences and the surface divergences. The bulk divergences, induced by the self-interaction within a single medium, are relatively simple and evaluated in the Appendix~\ref{ApB}. Here we concentrate on the surface divergences due to the interaction between two media, and investigate the behaviors of Casimir stresses near the surface as above, mainly about how the details of the inhomogeneity of the media varies the behaviors of surface divergences.

\par Notably, influences from the spatial distribution of the media and background on the divergences within the spherical systems have been recognized, and interesting arguments was displayed. For instance, the study about the divergent behaviors in the smooth and sharp background~\cite{bordag1999ground} can be stimulating. Typically, it is expected that the inhomogeneity of the media, especially at the surface, can reduce the surface divergences and the ambiguity in relevant physical quantities~\cite{milton2011hard,shayit2022vacuum}. However as shown by the solvable models above, the inhomogeneity does not guarantee well-defined behaviors at the surface. Better understandings on how the inhomogeneity works in softening the divergences are helpful. Rewrite Eq.~\eqref{eq.ge2} in terms of $q=(r-a)/a$ as
\begin{eqnarray}
\label{eq.TEeq2}
\bigg[\frac{d^2}{dq^2}-\frac{\mu'_t}{\mu_t}\frac{d}{dq}-\frac{\mu_t}{\mu_p}\frac{l(l+1)}{(1+q)^2}-\varepsilon_t\mu_t\zeta^2a^2\bigg]\tilde{e}_{\pm}(q)=0
,\quad
\end{eqnarray}
and the following forms for the permittivity and permeability of medium $i=1,2$ in the vicinity of the interface $r=a$ are used
\begin{subequations}
\label{eq.pp2}
\begin{eqnarray}
\varepsilon_{it(p)}(q)=\varepsilon_{it(p),0}\sum_{k=0}^{\infty}\alpha_{it(p),k}q^k,\ \alpha_{t(p),0}=1
,
\end{eqnarray}
\begin{eqnarray}
\mu_{it(p)}(q)=\mu_{it(p),0}\sum_{k=0}^{\infty}\beta_{it(p),k}q^k,\ \beta_{it(p),0}=1
.
\end{eqnarray}
\end{subequations}
With the permittivity and permeability in the form as Eq.~\eqref{eq.pp2}, Eq.~\eqref{eq.TEeq2} can be explicitly written as (with the label for the media $i=1,2$ ignored for brevity)
\begin{eqnarray}
\label{eq.Aq1}
&&
\bigg\{\frac{d^2}{dq^2}-\beta_{t,1}\frac{d}{dq}-\kappa^2
-\bigg(\frac{\sum\limits_{k=0}^{\infty}(k+1)\beta_{t,k+1}q^k}{\sum\limits_{k=0}^{\infty}\beta_{t,k}q^k}-\beta_{t,1}\bigg)\frac{d}{dq}
\nonumber\\
&&\quad
-k_l^2\bigg[\frac{\sum\limits_{k=0}^{\infty}\beta_{t,k}q^k}{\sum\limits_{k=0}^{\infty}\beta_{p,k}q^k}\frac{1}{(1+q)^2}-1\bigg]
\nonumber\\
&&\quad
-k_a^2\bigg[\sum_{k=0}^{\infty}\alpha_{t,k}q^k\sum_{k=0}^{\infty}\beta_{t,k}q^k-1\bigg]\bigg\}\tilde{e}_{\pm}(q)=0 ,
\end{eqnarray}
or in a simpler notation
\begin{eqnarray}
\label{eq.Aq2}
&&
\bigg(\frac{d^2}{dq^2}-\beta_{t,1}\frac{d}{dq}-\kappa^2
-\sum_{k=1}^{\infty}\rho_{1,k}q^k\frac{d}{dq}
-k_l^2\sum_{k=1}^{\infty}\rho_{2,k}q^k
\nonumber\\
&&\quad
-k_a^2\sum_{k=1}^{\infty}\rho_{3,k}q^k\bigg)\tilde{e}_{\pm}(q)=0 ,
\end{eqnarray}
in which $k_l^2=\gamma_{\mu}^2l(l+1)$, $k_a^2=n^2\zeta^2a^2$, $\kappa=\sqrt{k_l^2+k_a^2}$, $\gamma_{\mu}^2=\mu_{t,0}/\mu_{p,0}$ and $n^2=\varepsilon_{t,0}\mu_{t,0}$. Suppose $\tilde{e}_{\pm}(q)$ is of the form
\begin{eqnarray}
\tilde{e}_{\pm}(q)=e^{\zeta_{\pm}q}\sum_{k=0}^{\infty}c_{\pm,k}q^k,
\end{eqnarray}
with $c_{\pm,0}=1$ and
\begin{eqnarray}
\label{eq.zeta1}
\zeta_{\pm}=\frac{\beta_{t,1}\mp\sqrt{\beta^{2}_{t,1}+4\kappa^2}}{2},
\end{eqnarray}
then the equation to solve becomes
\begin{eqnarray}
\label{eq.reduced1}
&&
\bigg[\frac{d^2}{dq^2}+(2\zeta_{\pm}-\beta_{t,1})\frac{d}{dq}
-\sum_{k=1}^{\infty}\rho_{1,k}\zeta_{\pm}q^k
-\sum_{k=1}^{\infty}\rho_{1,k}q^k\frac{d}{dq}
\nonumber\\
&&\quad
-k_l^2\sum_{k=1}^{\infty}\rho_{2,k}q^k
-k_a^2\sum_{k=1}^{\infty}\rho_{3,k}q^k\bigg]\sum_{k=0}^{\infty}c_{\pm,k}q^k=0,
\end{eqnarray}
The few leading coefficients satisfy
\begin{eqnarray}
&&
2c_{\pm,2}+(2\zeta_{\pm}-\beta_{t,1})c_{\pm,1}=0,\ c_{\pm,3}=0,
\\
&&
2(2\zeta_{\pm}-\beta_{t,1})c_{\pm,2}-\rho_{1,1}c_{\pm,1}
=\rho_{1,1}\zeta_{\pm}
+k_l^2\rho_{2,1}
+k_a^2\rho_{3,1},\nonumber
\end{eqnarray}
which can be solved as
\begin{equation}
\label{eq.reduced1}
c_{\pm,1}
=
\frac{
\rho_{1,1}\zeta_{\pm}
+k_l^2\rho_{2,1}
+k_a^2\rho_{3,1}
}{
-(2\zeta_{\pm}-\beta_{t,1})^2-\rho_{1,1}
},\
c_{\pm,2}=\frac{\beta_{t,1}-2\zeta_{\pm}}{2}c_{\pm,1},
\end{equation}
and $\rho_{1,1}=2\beta_{t,2}-\beta_{t,1},\ \rho_{2,1}=\beta_{t,1}-\beta_{p,1}-2,\ \rho_{3,1}=\alpha_{t,1}+\beta_{t,1}$. For $k>3$, Eq.~\eqref{eq.reduced1} becomes
\begin{eqnarray}
\label{eq.reduced2}
&&
k(k-1)c_{\pm,k}+(k-1)(2\zeta_{\pm}-\beta_{t,1})c_{\pm,k-1}
\nonumber\\
&&
-\sum_{n=1}^{k-2}\rho_{1,n}\zeta_{\pm}c_{\pm,k-2-n}
-\sum_{n=1}^{k-2}(k-1-n)\rho_{1,n}c_{\pm,k-1-n}
\nonumber\\
&&\quad
-k_l^2\sum_{n=1}^{k-2}\rho_{2,n}c_{\pm,k-2-n}
-k_a^2\sum_{n=1}^{k-2}\rho_{3,n}c_{\pm,k-2-n}=0.\quad
\end{eqnarray}
So in principle the behaviors of $\tilde{e}_{\pm}$, and thus impacts on the behaviors of Casimir stresses, can be explicitly figured out with a little more algebra, one of which is about the ``softening'' of the divergences around the surface due to the inhomogeneity as mentioned above, and this ``soft wall'' effects can be thusly understood.

\par According to Eqs.~\eqref{eq.Aq1} and \eqref{eq.Aq2}, if we defined a matching between $\alpha_k,\beta_k$ and $q^k$, then as this matching extend to $\rho$s in Eq.~\eqref{eq.Aq2}. $\rho_{1,k}$ has the expression
\begin{eqnarray}
\rho_{1,k}=\sum_{\sum\limits_{q=0}^mn_q=k}
(-1)^{\sum\limits_{q=1}^mn_q}(n_0+1)\beta_{t,n_0+1}\prod_{q=1}^m\beta_{t,q},\quad
\end{eqnarray}
which shows that $\rho_{1,k}$ matches $q^{k+1}$. Similarly, we can see that, to the highest order of $q$, $\rho_{2,k}$ and $\rho_{3,k}$ match $q^k$. Further assume the matching between $k_l,k_a$ and $q$, then according to Eqs.~\eqref{eq.reduced1} and \eqref{eq.reduced2}, to the highest order of $q$, $c_{\pm,k}$ matches $q^k$. In the close vicinity of the interface, the following substitutions are used
\begin{eqnarray}
k_l^2q^2\rightarrow\frac{\mu_{t,0}}{\mu_{p,0}}k^2,\ \sum_{l=1}\nu q^2\rightarrow\int^\infty_0dkk,
\end{eqnarray}
then $\Delta T_{E;rr}$ and $\Delta T_{E;tt}$, as the instance without losing any generity, in Eq.~\eqref{eq.DT0} near the interface $r=a$ has the form
\begin{subequations}
\label{eq.TEnear}
\begin{eqnarray}
&&
\Delta T_{E;rr}(\mathbf{r})
=
\int_{0}^{\infty}\frac{-dkk}{4\pi^2 a^4q^4(1+q)^2}\int^{\infty}_0\frac{d\zeta}{\hat{d}_{1+}-\hat{d}_{1-}}
\nonumber\\
&&\quad
\times
\frac{\hat{d}_{1+}-y_{\mu}\hat{d}_{2+}}{y_{\mu}\hat{d}_{2+}-\hat{d}_{1-}}
\bigg\{
\frac{1}{\sum\limits_{i=0}^{\infty}\hat{\beta}_{1p,i}}\bigg(\sum_{i=1}^{\infty}i\hat{c}_{1-,i}\bigg)^2
-
\bigg(\sum_{i=0}^{\infty}\hat{c}_{1-,i}\bigg)^2
\nonumber\\
&&
\quad\quad
\times
\bigg[
n_1^2\zeta^2\sum\limits_{i=0}^{\infty}\hat{\alpha}_{1t,i}
+
\frac{\gamma_{1\mu}^2k^2}{(1+q)^2\sum\limits_{i=0}^{\infty}\hat{\beta}_{1p,i}}
\bigg]
\bigg\}e^{-2\hat{\zeta}_{1-}}
,
\end{eqnarray}
\begin{eqnarray}
&&
\Delta T_{E;tt}(\mathbf{r})=\int_{0}^{\infty}\frac{-dkk}{4\pi^2 a^4q^4(1+q)^2}\int^{\infty}_0\frac{d\zeta}{\hat{d}_{1+}-\hat{d}_{1-}}
\nonumber\\
&&\quad
\times
\frac{\hat{d}_{1+}-y_{\mu}\hat{d}_{2+}}{y_{\mu}\hat{d}_{2+}-\hat{d}_{1-}}
\frac{\gamma_{1\mu}^2k^2e^{-2\hat{\zeta}_{1-}}}{(1+q)^2\sum\limits_{i=0}^{\infty}\hat{\beta}_{1p,i}}
\bigg(\sum_{i=0}^{\infty}\hat{c}_{1-,i}\bigg)^2
,\quad\quad
\end{eqnarray}
\end{subequations}
where $y_{\mu}=\mu_{1t,0}/\mu_{2t,0}$, $\gamma_{1\mu}^2=\mu_{t,0}/\mu_{p,0}$, $n_1=\sqrt{\varepsilon_{1t,0}\mu_{1t,0}}$, $\hat{d}_{i\pm}=\hat{\zeta}_{i\pm}+\hat{c}_{i\pm,1}$, $\hat{\beta}_{k}$ is the fully ``matched'' $\beta_k$, i.e. $\hat{\beta}_k=\beta_kq^k$. $\hat{\zeta}_\pm$ has the same form as in Eq.~\eqref{eq.zeta1}, except for replacing $\beta$ with $\hat{\beta}$ and $\kappa_1=\sqrt{\gamma_{1\mu}^2k^2+n_1^2\zeta^2}$. Similarly, $\hat{c}_{i\pm,k}$ has the same form as $c_{i\pm,k}$ with $\alpha,\ \beta,\ \zeta$ replaced by $\hat{\alpha},\ \hat{\beta},\ \hat{\zeta}$. For example, $\hat{c}_{\pm,k}$ is
\begin{equation}
\hat{c}_{\pm,1}
=
-\frac{\hat{\rho}_{1,1}\hat{\zeta}_{\pm}
+\gamma_{\mu}k^2\hat{\rho}_{2,1}
+n_1^2\zeta^2\hat{\rho}_{3,1}}{(2\hat{\zeta}_{\pm}-\hat{\beta}_{t,1})^2+\hat{\rho}_{1,1}},
\end{equation}
in which we use the fact that $\rho_{1,k}$ matches $q^{k+1}$. For the ``hard wall'' case, where the two media are both of no inhomogeneity at all near their interface, according to Eq.~\eqref{eq.TEnear}, we see the highest order of divergences is $O(q^{-4})$, and just the corresponding planar stresses depending only on values of permitivities and permeabilities at the surface as before. Casimir stresses of the hard-wall spherical system also contain softer divergent terms due to the spherical geometry as shown in Eq.~\eqref{eq.vicinity}. When the contact between two media is ``softer'', that is, the media are inhomogeneous but their permittivities and permeabilities are equal at the interface, each contribution proportional to $\alpha$s, $\beta$s or their products will be softened by an order equal to the corresponding matching order defined above. Remarkably, one may gain an impression that only the media properties close to the surface matters, which is definitely not true. The geometry is vital. Besides the spherical modes used for description and the fact that the dependence on the radius of the ball is hidden in $q$, there is equivalent ``inhomogeneous'' terms in wave equation, such as the last two $q$-dependent terms in Eq.~\eqref{eq.equation02}. Needless to say the expressions of properties, as in Eq.~\eqref{eq.pp2}, also relies on the radius implicitly. Additionally, the spherical geometry and the inhomogeneity of media introduce great complexities to the behaviors of Casimir stresses, both near and far away from the surface, but does not cause higher order divergences than $O(q^{-4})$.

\par Furthermore, the inhomogeneity of media can further soften the surface divergence. Therefore, the number of divergent terms in the Casimir stresses above is actually finite. But it does not mean these divergences can be renormalized without any ambiguity, since within the macroscopic theory framework, it is still not clear what are the proper criteria to uniquely fix a well-defined (finite stresses) and stable (net forces zero) interface. Nevertheless, according to Eq.~\eqref{eq.TEnear}, the general condition for a ``well-defined'' surface (both the $\Delta T_{rr}$ and $\Delta T_{tt}$) is, in a sense, available when the zeroth, first and second order of derivatives of the permittivity and permeability for two touching media are equal at their interface. However, in this case, the interaction Casimir stresses in the whole system are zero. So this condition is actually not the well-defined interface condition for two contacting media, but the condition for how two media join together becoming ``one'' single medium. We evidently see that it is generally hopeless to achieve a nontrivial well-defined or stable surface in the sense of Casimir interaction from the macroscopic theory, even the inhomogeneity is included. The microscopic theory or macroscopic theory properly counting the microscopic effects may be requisite to overcome this paradox.

\section*{Conclusions}
\par In this work, we investigate Casimir stresses within the concentric inhomogeneous, as well as homogeneous, spherical system, consisting of a ball immersed in a background. Various relevant factors, such as the anisotropy, the refractive difference etc, are included. Impacts from the inhomogeneity are explored. To this aim, we firstly demonstrate the homogeneous case, where we explicitly show the leading divergent behaviors of the interaction Casimir stresses near the surface. As expected, corresponding planar results are derived, since the local interaction between two media close to the surface dominates. This holds true for both homogeneous and inhomogeneous cases. Furthermore, the influences of the spherical geometry introduce significant corrections to the divergences, especially for the radial components of the Casimir stress which are zero for the homogeneous planar configuration, and a $O(q^{-3})$ divergence depending on the radius of the ball is seen. The analytical expresses are compared numerically with the full results, and the consistency is justified. Numerical evaluations are also employed to show the general behaviors of Casimir stresses. Although considerable complexities are introduced by factors, such as anisotropic and refractive indices, they do not play key role in the divergent difficulty at the interface.

\par In the attempt of outlining the behaviors of surface divergences and throwing some light on the physical origin, we consider solvable inhomogeneous models and the general case near the surface. Since the inhomogeneity modifies the wave form of the spherical mode vastly, its impacts on the interaction, as well as the bulk (see examples in Appendix~\ref{ApB}), Casimir stresses are prominent everywhere, as shown in Fig.~\ref{fig:A3}. In the vicinity of the surface, as mentioned, there are divergences the same as the planar case from the local interaction. Since only related to the infinitesimal separation between interacting media and values of permitivities and permeabilities at the surface, properties of the media in the microscopic scale around the surface shall be essential, and the inhomogeneity of the media does not contribute to this kind of divergences. It is further demonstrated that there are secondary divergences due to the spherical geometry, and the inhomogeneity can soften the surface divergence. However in general the well-defined or stable nontrivial surface for the Casimir interaction is not available in the framework of macroscopic theory. But forming the trivial surface is possible, which requires the zeroth, first and second order of derivatives of the permittivity and permeability of two media at the surface being equal. Then the interaction Casimir stresses are zero everywhere, and two touching media in the spherical configuration with this condition satisfied, may thus be regarded as one single medium from the perspective of Casimir interaction.

\par To describe the Casimir interaction and extract relevant physical quantities and observables, the divergences should be properly handled, or even systematically renormalized, and many efforts have been devoted to come up with a usable renormalization scheme with diverse methods utilized and excellent results obtained. For instance, regarding the spatially varying potential as dynamical variables, the renormalization of coupling constants were carried out~\cite{mazzitelli2011boundary,fulling2012energy}, with the renormalizable model for quantum scalar field~\cite{mazzitelli2011boundary} and divergences near a soft wall weaker than that around the infinitely high potential wall found~\cite{fulling2012energy}. By imposing some physical constrains, such as the conservation, covariance, trace equation etc, on the ambiguous terms, Milton et. al.~\cite{milton2016stress} proposed a ``renormalization'' procedure for the scalar field with a spatailly varying in one direction in the palnar system. For the spherical system, Ref.~\cite{avni2018casimir} suggested calculating the difference in the radial stress in- and outside the surface of the ball at $a+\Delta$ and $a-\Delta$, and counting the finite term as $\Delta\rightarrow0$ as the results. Yet the divergences, especially surface divergences, can still exist though maybe weaker~\cite{mazzitelli2011boundary,fulling2012energy,milton2016stress}. The scheme of Ref.~\cite{avni2018casimir} is also in discussion~\cite{milton2020self}. A general renormalization approach to deal with the divergences when evaluating the Casimir interaction is still to be explored.

\par In this work, we mostly explore the behaviors of divergences, and according to our results, the small distance between the interacting media across the interface is a main source of the surface divergence. It should be desirable that more physics can be unveiled via going beyond the macroscopic theory~\cite{mahanty1977colloid,griniasty2017casimir,griniasty2017casimir} and fully taking the microscopic features of the media near the surface and the properties of field in small scales~\cite{yang2019general} into account. In our future work, we will concentrate on the renormalization of surface divergence by incorporating the microscopic aspects of the surface. On the other hand, as we have shown above, the significance of the global geometry should never be underestimated, even for understanding divergences attached to the surface. The transition from the microscopic scale near the surface to the usual macroscopic form in the bulk for the spherical geometry and its impacts on the the global aspect of the Casimir interaction would be our primary concern as well. For both the investigation on the microscopic scale and the transition to the bulk, the influences from the inhomogeneity within the theory framework should never be overlooked.

\appendix
\section{Casimir stresses in the homogeneous planar configuration}
\label{ApA}
\par In the work, the planar configuration are employed to make comparisons with corresponding spherical results repeatedly. For the sake of convenience and self-containment, here we provide the brief evaluation and discussion for the planar system. The model and method utilized are both in line with those for the spherical configuration, in the hope that not only the results are compared easily, but also the similarity and difference of between the theoretical structures spherical and planar can be spotted explicitly. Similar method is also used in Ref.~\cite{parashar2018quantum}, where readers interested in more details about the method and the Casimir stresses in the inhomogeneous planar configurations are referred to.

\par For planar cases where the media is isotropic parallel to the interface and anisotropic norm to the interface (the interface here is assumed to be the plane $z=0$), the Green's dyadic $\bm{\Gamma}$ has a form distinguishing from those in Eq.~\eqref{eq.spGamma1} and \eqref{eq.spGamma2}
\begin{subequations}
\begin{equation}
\bm{\Gamma}_\zeta(\mathbf{r},\mathbf{r}')=
\int\frac{d^2k}{(2\pi)^2}e^{i\mathbf{k}\cdot(\mathbf{r}_\parallel-\mathbf{r}'_\parallel)}
\mathbf{g}_{\zeta,\mathbf{k}}(z,z'),
\end{equation}
\begin{equation}
\mathbf{g}_{\zeta,\mathbf{k}}(z,z')=
\left[
  \begin{array}{ccc}
    \frac{\partial_z\partial_{z'}}{\varepsilon_t\varepsilon'_t}g^H_{\zeta,\mathbf{k}} &  & \frac{ik\partial_z}{\varepsilon_t\varepsilon'_p}g^H_{\zeta,\mathbf{k}} \\
     & -\zeta^2g^E_{\zeta,\mathbf{k}} &  \\
    -\frac{ik\partial_{z'}}{\varepsilon_p\varepsilon'_t}g^H_{\zeta,\mathbf{k}} &  & \frac{k^2}{\varepsilon_p\varepsilon'_p}g^H_{\zeta,\mathbf{k}} \\
  \end{array}
\right],
\end{equation}
where $\varepsilon_t$ and $\varepsilon_p$ ($\mu_t$ and $\mu_p$) are respectively the components of the permittivity (permeability) parallel and norm to the interface,  $g^E_{\zeta,\mathbf{k}}$ and $g^H_{\zeta,\mathbf{k}}$ satisfy
\begin{equation}
\bigg[\partial_z\frac{1}{\mu_t}\partial_z-\varepsilon_t\zeta^2-\frac{k^2}{\mu_p}\bigg]
g^E_{\zeta,\mathbf{k}}(z,z')=\delta(z-z'),
\end{equation}
\begin{equation}
\bigg[\partial_z\frac{1}{\varepsilon_t}\partial_z-\mu_t\zeta^2-\frac{k^2}{\varepsilon_p}\bigg]
g^H_{\zeta,\mathbf{k}}(z,z')=\delta(z-z').
\end{equation}
\end{subequations}
So the TE Casimir stress tensors can be derived similarly as in the spherical case
\begin{subequations}
\label{eq.TEstressP}
\begin{equation}
T_{E;zz}=-\int\frac{d\zeta d^2k}{(2\pi)^3}\frac{1}{2}\bigg(\frac{\partial_z\partial_{z'}}{\mu'_t}g^E_{\zeta,\mathbf{k}}
-\varepsilon_t\zeta^2g^E_{\zeta,\mathbf{k}}-\frac{k^2}{\mu_p}g^E_{\zeta,\mathbf{k}}\bigg),
\end{equation}
\begin{eqnarray}
&&
T_{E;xx}=-\int\frac{d\zeta d^2k}{(2\pi)^3}\frac{1}{2}\bigg[\frac{k_y^2-k_x^2}{k^2}\bigg(\frac{\partial_z\partial_{z'}}{\mu'_t}g^E_{\zeta,\mathbf{k}}
+\varepsilon_t\zeta^2g^E_{\zeta,\mathbf{k}}\bigg)
\nonumber\\
&&\quad\quad
+\frac{k^2}{\mu_p}g^E_{\zeta,\mathbf{k}}\bigg],
\end{eqnarray}
\begin{eqnarray}
&&
T_{E;yy}=-\int\frac{d\zeta d^2k}{(2\pi)^3}\frac{1}{2}\bigg[\frac{k_x^2-k_y^2}{k^2}\bigg(\frac{\partial_z\partial_{z'}}{\mu'_t}g^E_{\zeta,\mathbf{k}}
+\varepsilon_t\zeta^2g^E_{\zeta,\mathbf{k}}\bigg)
\nonumber\\
&&\quad\quad
+\frac{k^2}{\mu_p}g^E_{\zeta,\mathbf{k}}\bigg],
\end{eqnarray}
\end{subequations}
their TM counterparts can be obtained by making the substitution $E\rightarrow H,\ \varepsilon\leftrightarrow\mu$. Write
\begin{equation}
\bigg[\partial_z\frac{1}{\mu_t}\partial_z-\varepsilon_t\zeta^2-\frac{k^2}{\mu_p}\bigg]e_{\pm}(z)=0,
\end{equation}
with proper boundary satisfied, i.e. $z\rightarrow\pm\infty,\ e_{\pm}(z)\rightarrow0$, then
\begin{equation}
\label{eq.gEp}
g^E(z,z')=\frac{e_+(z_>)e_-(z_<)}{W^E},\ W^E=\frac{e'_+e_--e_+e'_-}{\mu_t}.
\end{equation}
If we have medium 1 in $z<0$ and medium 2 in $z>0$, then $e_{\pm}(z)$ are solved as
\begin{subequations}
\begin{equation}
e_+(z)=
\left\{
  \begin{array}{cc}
    e_{2+}(z), & z>0, \\
    A_ee_{1+}(z)+B_ee_{1-}(z), & z<0, \\
  \end{array}
\right.
\end{equation}
\begin{equation}
e_-(z)=
\left\{
  \begin{array}{cc}
    C_ee_{2+}(z)+D_ee_{2-}(z), & z>0, \\
    e_{1-}(z), & z<0, \\
  \end{array}
\right.
\end{equation}
with the coefficients and the generalized Wronskian being
\begin{equation}
A_e=\frac{[e_{2+},e_{1-}]_{\mu}(0)}{W^E_1},\ B_e=\frac{[e_{1+},e_{2+}]_{\mu}(0)}{W^E_1},
\end{equation}
\begin{equation}
C_e=\frac{[e_{1-},e_{2-}]_{\mu}(0)}{W^E_2},\ D_e=\frac{[e_{2+},e_{1-}]_{\mu}(0)}{W^E_2},
\end{equation}
\end{subequations}
and $W^E=[e_{2+},e_{1-}]_{\mu}(0)$. In terms of $e_{i\pm}$, the stresses in each region can be written as follows: for $z<0$
\begin{subequations}
\begin{equation}
T_{E;zz}=-\int\frac{d\zeta d^2k}{(2\pi)^3}\frac{B_e}{2W^E}\bigg[\frac{e_{1-}^{\prime2}(z)}{\mu_{1t}}
-\bigg(\varepsilon_{1t}\zeta^2+\frac{k^2}{\mu_{1p}}\bigg)e_{1-}^2\bigg],
\end{equation}
\begin{eqnarray}
&&
T_{E;xx}=-\int\frac{d\zeta d^2k}{(2\pi)^3}\frac{B_e}{2W^E}\bigg[\frac{k_y^2-k_x^2}{k^2}\bigg(\frac{e_{1-}^{\prime2}(z)}{\mu_{1t}}
\nonumber\\
&&\quad\quad
+\varepsilon_{1t}\zeta^2e_{1-}^2(z)\bigg)+\frac{k^2}{\mu_{1p}}e_{1-}^2(z)\bigg],
\end{eqnarray}
\begin{eqnarray}
&&
T_{E;yy}=-\int\frac{d\zeta d^2k}{(2\pi)^3}\frac{B_e}{2W^E}\bigg[\frac{k_x^2-k_y^2}{k^2}\bigg(\frac{e_{1-}^{\prime2}(z)}{\mu_{1t}}
\nonumber\\
&&\quad\quad
+\varepsilon_{1t}\zeta^2e_{1-}^2(z)\bigg)+\frac{k^2}{\mu_{1p}}e_{1-}^2(z)\bigg],
\end{eqnarray}
\end{subequations}
while for $z>0$
\begin{subequations}
\begin{equation}
T_{E;zz}=-\int\frac{d\zeta d^2k}{(2\pi)^3}\frac{C_e}{2W^E}\bigg[\frac{e_{2+}^{\prime2}(z)}{\mu_{2t}}
-\bigg(\varepsilon_{2t}\zeta^2+\frac{k^2}{\mu_{2p}}\bigg)e_{2+}^2(z)\bigg],
\end{equation}
\begin{eqnarray}
&&
T_{E;xx}=-\int\frac{d\zeta d^2k}{(2\pi)^3}\frac{C_e}{2W^E}\bigg[\frac{k_y^2-k_x^2}{k^2}\bigg(\frac{e_{2+}^{\prime2}(z)}{\mu_{2t}}
\nonumber\\
&&\quad\quad
+\varepsilon_{2t}\zeta^2e_{2+}^2(z)\bigg)+\frac{k^2}{\mu_{2p}}e_{2+}^2(z)\bigg],
\end{eqnarray}
\begin{eqnarray}
&&
T_{E;yy}=-\int\frac{d\zeta d^2k}{(2\pi)^3}\frac{C_e}{2W^E}\bigg[\frac{k_x^2-k_y^2}{k^2}\bigg(\frac{e_{2+}^{\prime2}(z)}{\mu_{2t}}
\nonumber\\
&&\quad\quad
+\varepsilon_{2t}\zeta^2e_{2+}^2(z)\bigg)+\frac{k^2}{\mu_{2p}}e_{2+}^2(z)\bigg].
\end{eqnarray}
\end{subequations}

\par When the media on both sides of their interface are homogeneous, then
\begin{equation}
e_{i\pm}(z)=e^{\mp\kappa_iz},\ W^E_i=-\frac{2\kappa_i}{\mu_{it}},
\end{equation}
with parameters being $\kappa_i=\sqrt{n_i^2\zeta^2+\gamma_{i\mu}^2k^2},\ n_i=\sqrt{\varepsilon_{it}\mu_{it}},\ \gamma_{i\mu}^2=\frac{\mu_{it}}{\mu_{ip}}$. Obviously $T_{E;zz}$ now is zero everywhere. The transverse components of the stress tensor in $z<0$ region are
\begin{subequations}
\label{eq.pltr}
\begin{equation}
T_{E;xx}=T_{E;yy}=
\frac{1}{4\pi^2z^4n_1}\frac{-3}{16\gamma_{1\mu}^2}\int^{\pi/2}_0d\theta\frac{\kappa_2y_{\mu}-1}{\kappa_2y_\mu+1}
\sin^3\theta
,
\end{equation}
with $\kappa_2$ changed to $\kappa_2=\sqrt{n_{21}^2\cos^2\theta+\frac{\gamma_{2\mu}^2}{\gamma_{1\mu}^2}\sin^2\theta}$ and $y_{\mu}=\frac{\mu_{1t}}{\mu_{2t}}$, while for $z>0$
\begin{equation}
T_{E;xx}=T_{E;yy}
=
\frac{1}{4\pi^2n_2z^4}\frac{-3}{16\gamma_{2\mu}^2}\int^{\pi/2}_0d\theta\frac{\kappa_1-y_\mu}{y_\mu+\kappa_1}
\sin^3\theta
,
\end{equation}
\end{subequations}
with $\kappa_1$ changed to $\kappa_1=\sqrt{n_{12}^2\cos^2\theta+\frac{\gamma_{1\mu}^2}{\gamma_{2\mu}^2}\sin^2\theta}$.

\section{Bulk Casimir stresses}
\label{ApB}
\par This work focus on the interaction contributions to the Casimir stresses. Here we briefly consider the bulk contributions for the models involved in the main text.

\par For the homogeneous background with the permittivity and permeability of the same form as in Eq.~\eqref{eq.spepsilon}, we have
\begin{equation}
\label{eq.es}
\tilde{e}_{+}(r)=e_{\nu_{\mu}}(\kappa r),\ \tilde{e}_{-}(r)=s_{\nu_{\mu}}(\kappa r),
\end{equation}
where $\kappa^2=\varepsilon_{t}\mu_{t}\zeta^2=n^2\zeta^2$, $\nu_{\mu}=\sqrt{\gamma_{\mu}^2l(l+1)+1/4}$, $\gamma_{\mu}^2=\mu_{t}/\mu_{p}$, $e_n(x)$ and $s_n(x)$ are the modified Ricatti-Bessel functions defined as
\begin{equation}
\label{eq.es}
e_n(x)=\sqrt{\frac{2x}{\pi}}K_{n}(x),\ s_n(x)=\sqrt{\frac{\pi x}{2}}I_{n}(x).
\end{equation}
and the bulk stresses are then
\begin{subequations}
\begin{eqnarray}
&&
T_{E;rr}
=
\sum_{l=1}^{\infty}\frac{\nu P_l(\cos\alpha)}{4\pi r^4}\int^\infty_0\frac{dxx}{n\pi}
\cos(x\tau')
\bigg\{
e'_{\nu_{\mu}}(x)s'_{\nu_{\mu}}(x)
\nonumber\\
&&\quad
-\bigg[\frac{\nu_{\mu}^2-1/4}{x^2}+1\bigg]e_{\nu_{\mu}}(x)s_{\nu_{\mu}}(x)
\bigg\},\ \tau'=\tau/nr,
\end{eqnarray}
\begin{eqnarray}
&&
T_{E;tt}
=
\sum_{l=1}^{\infty}\frac{\nu P_l(\cos\alpha)}{4\pi r^4}\int^\infty_0\frac{dxx}{n\pi}\cos(x\tau')
\frac{\nu_{\mu}^2-1/4}{x^2}
\nonumber\\
&&\quad
\times
e_{\nu_{\mu}}(x)s_{\nu_{\mu}}(x)
,
\end{eqnarray}
\end{subequations}
where the temporal ($\tau\rightarrow0$) and spatial (here as the angle between two splitting points $\alpha\rightarrow0$) point-splitting regulators are introduced. When the background is isotropic, then $\nu_{\mu}=\nu=l+1/2$, and the stresses are
\begin{subequations}
\begin{eqnarray}
&&
T_{E;rr}
=
\sum_{l=1}^{\infty}\frac{\nu P_l(\cos\alpha)}{4\pi r^4}\int^\infty_0\frac{dxx}{n\pi}
\cos(x\tau')
\bigg\{
e'_{\nu}(x)s'_{\nu}(x)
\nonumber\\
&&\quad
-e_{\nu}(x)s_{\nu}''(x)
\bigg\},
\end{eqnarray}
\begin{eqnarray}
&&
T_{E;tt}
=
\sum_{l=1}^{\infty}\frac{\nu P_l(\cos\alpha)}{4\pi r^4}\int^\infty_0\frac{dxx}{n\pi}\cos(x\tau')
\bigg\{
e_{\nu}(x)s_{\nu}''(x)
\nonumber\\
&&\quad
-
e_{\nu}(x)s_{\nu}(x)
\bigg\}
.
\end{eqnarray}
\end{subequations}
By utilizing the following formula~\cite{klich1999casimir},
\begin{eqnarray}
\sum_{l=1}^{\infty}\nu P_l(\cos\alpha)e_{\nu}(x)s_{\nu}(y)
=\frac{xy}{2\rho}e^{-\rho}-\frac{1}{2}e^{-x}\sinh(y)
,\quad
\end{eqnarray}
where $\rho=\sqrt{x^2+y^2-2xy\cos\alpha}$, the stresses can be evaluated as
which means
\begin{subequations}
\label{eq.TsES}
\begin{eqnarray}
&&
T_{E;rr}
=
\frac{-1}{8\pi^2nr^4\tau^{'2}}
+
\frac{1}{8\pi^2nr^4}
\frac{4+\tau^{'2}}{(\alpha^2+\tau^{'2})^2}
,
\end{eqnarray}
\begin{eqnarray}
&&
T_{E;tt}
=
\frac{-1}{4\pi^2nr^4}
\frac{2\alpha^2+\alpha^2\tau^{'2}-2\tau^{'2}}{(\alpha^2+\tau^{'2})^3}.
\end{eqnarray}
\end{subequations}
As a comparison, consider the planar counterparts of the anisotropic homogeneous model above, that is, the uniform background with the permittivity and permeability being $\bm{\varepsilon}/\bm{\mu}=\varepsilon_t/\mu_t(\mathbf{\hat{x}\hat{x}}+\mathbf{\hat{y}\hat{y}})
+\varepsilon_p/\mu_p\mathbf{\hat{z}\hat{z}}$. Then substitute the following functions into Eqs.~\eqref{eq.gEp} and \eqref{eq.TEstressP}
\begin{eqnarray}
&&
e_{\pm}(z)=e^{\mp\kappa z},\ W^E=-\frac{2\kappa}{\mu_{t}},\ \kappa=\sqrt{n^2\zeta^2+\gamma_{\mu}^2k^2},
\nonumber\\
&&\quad
n=\sqrt{\varepsilon_t\mu_t},\ \gamma_{\mu}=\sqrt{\mu_t/\mu_p},
\end{eqnarray}
together with the temporal ($\tau\rightarrow0$) and spatial ($\bm{\delta}\rightarrow\bm{0}$) point-splitting regulators, the bulk stresses can be expressed as
\begin{subequations}
\label{eq.TsEP}
\begin{eqnarray}
&&
T_{E;zz}
=
\frac{2}{4\pi^2n\gamma_{\mu}^2d^4},
\end{eqnarray}
\begin{eqnarray}
&&
T_{E;xx}
=
\frac{2}{4\pi^2n\gamma_{\mu}^2d^4}
-
\frac{8\delta_x^{'2}}{4\pi^2n\gamma_{\mu}^2d^6}
,
\end{eqnarray}
\begin{eqnarray}
&&
T_{E;yy}
=
\frac{2}{4\pi^2n\gamma_{\mu}^2d^4}
-
\frac{8\delta_y^{'2}}{4\pi^2n\gamma_{\mu}^2d^6}
,
\end{eqnarray}
\end{subequations}
where $\tau'=\frac{\tau}{n},\bm{\delta}=\frac{\bm{\delta}'}{\gamma_{\mu}},d=\sqrt{\tau^{'2}+\bm{\delta}^{'2}}$. Since for different geometries the explicit form of the spatial regulator used should be specifically chosen to get a simple formula, they give different expressions for the stresses. As shown above, for the isotropic homogeneous background, the stresses show different forms in Eqs.~\eqref{eq.TsES} and \eqref{eq.TsEP}. If only the temporal regulator is used, i.e. $\delta=0$ in Eq.~\eqref{eq.TsES} and $\bm{\delta}=\bm{0}$ in Eq.~\eqref{eq.TsEP}, then the stresses are the same in each direction. So in the following arguments, we keep only the temporal point-splitting to get rid of the extra complexities of the spatial regulator.

For the anisotropic case, $\gamma_{\mu}\neq1$, we employ UAE to demonstrate the leading behaviors of the stresses as
\begin{subequations}
\label{eq.TsESA}
\begin{eqnarray}
\label{eq.TsESA1}
&&
T_{E;rr}
\approx
\sum_{l=1}^{\infty}\frac{\nu}{8\pi^2nr^4}\int^\infty_0dx
\cos(x\tau')
\bigg[
-2(\nu_{\mu}^2+x^2)^{\frac{1}{2}}
\nonumber\\
&&\quad
+
\frac{1}{4(\nu_{\mu}^2+x^2)^{\frac{1}{2}}}
+
\frac{\nu_{\mu}^4}{4(\nu_{\mu}^2+x^2)^{\frac{5}{2}}}
\bigg]
\nonumber\\
&=&
\frac{4}{8\pi^2n\gamma_{\mu}^2r^4\tau^{'4}}
+
\frac{1}{8\pi^2n\gamma_{\mu}^2r^4\tau^{'2}}
\frac{11}{12}
,
\end{eqnarray}
\begin{eqnarray}
\label{eq.TsESA2}
&&
T_{E;tt}
\approx
\sum_{l=1}^{\infty}\frac{\nu}{8\pi^2nr^4}\int^\infty_0dx
\cos(x\tau')
\bigg\{
\frac{\nu_{\mu}^2}{(\nu_{\mu}^2+x^2)^{\frac{1}{2}}}
\nonumber\\
&&\quad
-
\frac{1}{4(\nu_{\mu}^2+x^2)^{\frac{1}{2}}}
+
\frac{\nu_{\mu}^2}{8(\nu_{\mu}^2+x^2)^{\frac{3}{2}}}
\nonumber\\
&&\quad
-
\frac{3\nu_{\mu}^4}{4(\nu_{\mu}^2+x^2)^{\frac{5}{2}}}
+
\frac{5\nu_{\mu}^6}{8(\nu_{\mu}^2+x^2)^{\frac{7}{2}}}
\bigg\}
=
\frac{4}{8\pi^2n\gamma_{\mu}^2r^4\tau^{'4}}
,\quad\quad
\end{eqnarray}
\end{subequations}
where only the temporal regulator is used. The stresses, shown in Eq.~\eqref{eq.TsESA}, has the same leading divergences everywhere as the planar case in Eq.~\eqref{eq.TsEP}. The radial stress component in Eq.~\eqref{eq.TsESA1} contains a softer divergent term, which varies with the distance from the origin as $\sim r^{-2}$, signifying the convergence to the planar case as the curvature vanishes.

\par For inhomogeneous case, suppose the background has the properties
\begin{eqnarray}
\varepsilon_{t}(\mu_{t})=\frac{\varepsilon_{t0}(\mu_{t0})}{r}a,\ \varepsilon_{p}(\mu_{p})=\frac{\varepsilon_{p0}(\mu_{p0})}{r}a,
\end{eqnarray}
then $\tilde{e}_{\pm}$ can be solved as
\begin{eqnarray}
\tilde{e}_{\pm;\zeta,l}(r)=r^{\mp\xi_{\mu}},\ \xi_{\mu}=\sqrt{\nu_{\mu}^2-1/4+\varepsilon_{t0}\mu_{t0}\zeta^2a^2},\quad\quad
\end{eqnarray}
where $\nu_{\mu}=\sqrt{\gamma_{\mu}^2l(l+1)+1/4}$ and $\gamma_{\mu}^2=\mu_{t0}/\mu_{p0}$, and we have the stresses regularized by the temporal point-splitting $\tau$
\begin{eqnarray}
&&
T_{E;rr}=T_{E;tt}
=
\frac{2}{4\pi^2n\gamma_{\mu}^2r^3a}\frac{1}{\tau^{'4}}
,\ \tau'=\tau/na,
\end{eqnarray}
where $n=\sqrt{\varepsilon_{t0}\mu_{t0}}$. For another inhomogeneous case with
\begin{eqnarray}
\varepsilon_t(\mu_t)=\varepsilon_{t0}(\mu_{t0}),\ \varepsilon_p(\mu_p)=\frac{\varepsilon_{p0}(\mu_{p0})}{r^2}a^2,
\end{eqnarray}
then $\tilde{e}_{\pm}$ can be solved as
\begin{eqnarray}
\tilde{e}_{\pm;\zeta,l}(r)=e^{\mp\xi_{\mu}r},\ \xi_{\mu}=\sqrt{\frac{\nu_{\mu}-1/4}{a^2}+\varepsilon_{t0}\mu_{t0}\zeta^2},\quad
\end{eqnarray}
where $\nu_{\mu}=\sqrt{\gamma_{\mu}^2l(l+1)+1/4}$ and $\gamma_{\mu}^2=\mu_{t0}/\mu_{p0}$, and the stresses are
\begin{eqnarray}
T_{E;rr}
=
\frac{2}{4\pi^2n\gamma_{\mu}^2r^2a^2}\frac{1}{\tau^{'4}},\
T_{E;tt}
=
\frac{2}{4\pi^2n\gamma_{\mu}^2r^4}\frac{1}{\tau^{'4}}.\quad
\end{eqnarray}

\par As shown in the homogeneous cases, we can also decompose the electromagnetic field into modes of either planar wave (labeled with the wavevector of the plane wave $\mathbf{k}$ as in Appendix \ref{ApA}) or spherical wave (labeled by the angular index $l$) form, and for the isotropic homogeneous case, different structures of the modes combined with corresponding waveform for each mode lead us to the same results. In the inhomogeneous cases above, the structure of the modes does not change, but the waveform for each mode is significantly modified by the inhomogeneity of the media, resulting in significant variations of the stresses. But the divergences are not softened.

\begin{acknowledgments}
Y.L. was supported by the National Natural Science Foundation of China (grant number 12304396). Y.L. thanks the Terahertz Physics and Devices Group, Nanchang University for the strong computational facility support. We gratefully acknowledge constructive discussions with Dr. K.A.Milton, Y.Zhuo, and W.Q.Zhou.
\end{acknowledgments}

\bibliography{refbib}

\end{document}